\documentclass[12pt]{article}

\usepackage{epsfig}
\usepackage{graphicx}
\usepackage{amssymb,amsmath}
\usepackage{hyperref}
\usepackage[nosort]{cite}
\hypersetup{
	colorlinks=true,
	linkcolor=blue,
	citecolor=red,
}

\begin{document}
	
\begin{center}
	\Large{\bf Ho\v{r}ava-Lifshitz $F(\bar{R})$ theories and the Swampland} \vspace{0.5cm}
	
	\large  H. Garc\'{\i}a-Compe\'an\footnote{e-mail address: {\tt
			compean@fis.cinvestav.mx}}, D. Mata-Pacheco\footnote{e-mail
		address: {\tt dmata@fis.cinvestav.mx}}, L. Zapata\footnote{e-mail address:{ \tt lzapata@fis.cinvestav.mx}}
	
	\vspace{0.3cm}
	
	{\small \em Departamento de F\'{\i}sica, Centro de
		Investigaci\'on y de Estudios Avanzados del IPN}\\
	{\small\em P.O. Box 14-740, CP. 07000, Ciudad de M\'exico, M\'exico}\\

	\vspace*{1.5cm}
\end{center}	

\begin{abstract}
	
	The compatibility between the de Sitter Swampland conjecture and Ho\v{r}ava--Lifshitz $F(\bar{R})$  theories with a flat FLRW metric is studied. We first study the standard $f(R)$ theories and show that the only way in which the dS conjecture can be made independent of $R$ is by considering a power law of the form $f(R)\sim R^{\gamma}$. The conjecture and the consistency of the theory puts restrictions on $\gamma$ to be greater but close to one. For $F(\bar{R})$ theories described by its two parameters $\lambda$ and $\mu$, we use the equations of motion to construct the function starting with an ansatz for the scale factor in the Jordan frame of the power law form. By performing a conformal transformation on the three metric to the Einstein frame, we can obtain an action of gravity plus a scalar field by relating the parameters of the theory. The non-projectable and projectable cases are studied and the differences are outlined. The obtained $F(\bar{R})$ function consists of terms of the form $\bar{R}^{\gamma}$ with the possibility of having negative power terms. The dS conjecture leads to inequalities for the $\lambda$ parameter; in both versions, it becomes restricted to be greater but close to $1/3$. We can also study the general case in which $\mu$ and $\lambda$ are considered as independent. The obtained $F$ function  has the same form as before. The consistency of the theory and the dS conjecture lead to a set of inequalities on both parameters that are studied numerically. In all cases,  $\lambda$ is restricted by $\mu$ around $1/3$, and we obtain $\lambda\to1/3$ if $\mu\to0$. We consider the $f(R)$ limit $\mu,\lambda \to 1$ and we obtain consistent results. Finally, we study the case of a constant Hubble parameter. The dS conjecture can be fulfilled by restricting the parameters of the theory; however, the constraint makes this compatibility exclusive to these kinds of theories.
\vskip 1truecm

\end{abstract}

\bigskip

\newpage

%%%%%%%%%%%%%%%%%%%%%%%%%%%%%%%%%%%%%%%%%%%%%%%%%%%%%%%%%%%%%%%%%%%%%%%%%%%%%%%%%%%%%%
%%%%%%%%%%%%%%%%%%%%%%%%%%%%%%%%%%%%%%%%%%%%%%%%%%%%%%%%%%%%%%%%%%%%%%%%%%%%%%%%%%%%%%
%%%%%%%%%%%%%%%%%%%%%%%%%%%%%%%%%%%%%%%%%%%%%%%%%%%%%%%%%%%%%%%%%%%%%%%%%%%%%%%%%%%%%%

\section{Introduction}
\label{S-Intro}
One fundamental aspect of any physical theory is the agreement that it must hold with experiment. However, it has been known from the beginning that any approach to a quantum theory of gravity would have severe issues with this aspect, and string theory is not an exception. Decades have passed since the first string theory was proposed. In that amount of time, the theory has been improved vastly, leading to a framework with a vast scenario of possible realizations.  One of the most controversial issues that string theory possesses in this regard is the choice of a vacuum for the theory, since the physical constants on any string theory come from vacuum expectation values for scalar fields. The choice of such a vacuum is fundamental. However, at present, no mechanism to choose dynamically such a vacuum exists. Instead, there is a very big number of possible vacua that can be compatible, the so-called Landscape of string theory. If we have a full quantum theory of gravity, any model constructed within it leads to an effective theory in the low energy limit. However, the opposite is not true in general , and thus the idea that it should be possible to distinguish an effective theory that can be completed to a correct quantum theory of gravity in the ultraviolet (UV) from those which cannot  is an important subject. The theories that can be completed in this sense are said to belong to the mentioned Landscape. In contrast, if an effective theory in the infrared (IR) cannot be completed in this sense, it is said to belong to the Swampland. In order to pursue conditions on an effective field theory containing gravity to be completed in the UV,  a set of Swampland conjectures have been proposed recently \cite{Vafa:2005ui,Brennan:2017rbf,Ooguri:2006in,Palti:2019pca,vanBeest:2021lhn,Grana:2021zvf}. They describe some behaviours found generically on string theoretic realizations that seem to be key elements of a quantum gravity theory. Although features derived from string theory models are used as guidelines, these conjectures are expected to be valid in a general sense independently of string theory.

There is a vast number of constructions in string theory aiming to describe a cosmological scenario compatible with observations. For example, from the inflationary point of view, string theory has many proposals for the inflaton field (see, for example, \cite{Baumann:2014nda}). However, it was found very difficult to obtain a dS background with a complete construction on the higher dimensional theory. This seeming impossibility led to the dS conjecture that now claims that it is not possible to obtain such a background \cite{Obied:2018sgi,Ooguri:2018wrx,Andriot:2018wzk,Agrawal:2018own,Roupec:2018mbn}. There are other Swampland conjectures in this sense, such as the Distance conjecture, the Transplanckian Censorship conjecture, etc., but in this article, we focus exclusively on the dS conjecture.  When we only have one scalar field coupled to gravity as standard in inflationary models, this conjecture is written as inequalities on the scalar field potential. However, it is known that the dS conjecture is not compatible with the standard cosmological picture of an inflationary universe driven by a scalar field that obeys the slow-roll conditions on General Relativity (GR) \cite{Garg:2018reu,Ben-Dayan:2018mhe,Kinney:2018nny}, since those conditions are expressed in terms of the scalar field potential and they are the opposite of the dS conjecture among other problems. There are various ways out of this inconsistency such as considering Warm Inflation \cite{Motaharfar:2018zyb}, Multi-field inflation~\cite{Achucarro:2018vey} or variations of the gravitational theory such as $f(R)$ theories~\cite{Artymowski:2019vfy,Benetti:2019smr,Elizalde:2022oej}  among other proposals \cite{Denef:2018etk,Trivedi:2020wxf,Yi:2018dhl,Brahma:2019kch,Trivedi:2021nss}. In this work, we are interested in studying the compatibility of this conjecture with other modifications of the gravitational theory known as Ho\v{r}ava--Lifshitz $F(\bar{R})$ theories.
 
It is known that GR must be the theory of gravity on the low-energy regime. String theory, of course, fulfills this condition. However, instead of modifying the theory of gravity at the fundamental level to discover its quantum description, there is also the possibility of improving GR to obtain such description in a closer way to the original theory. In this sense, inspired by the Lifshitz scaling of condensed matter systems, Petr Ho\v{r}ava proposed to modify the Einstein--Hilbert action by incorporating spatial higher-derivative terms compatible with an anisotropic scaling of space and time variables \cite{Horava:2009uw,Bertolami:2011ka,Sotiriou:2010wn,Kiritsis:2009sh,Mukohyama:2010xz}. These terms break the Lorentz symmetry in the UV but make the theory power counting renormalizable, and thus it represents a better candidate to a quantum UV completion of gravity. The IR limit of this theory was found to be troublesome, but it led to new relevant cosmological behaviours such as bouncing universes \cite{Calcagni:2009ar,Brandenberger:2009yt,Czuchry:2009hz} or the appearance of a dark matter term as a constant of integration \cite{Mukohyama:2009mz} to name a few. This theory has been generalized in a number of ways in order to improve the issues in the IR limit found in the original theory. For example, in \cite{Zhu:2011yu,Huang:2012ep}, an extra $U(1)$ field was added. A generalization of such a theory has also been proposed in the same way as $f(R)$ theories are a generalization of the standard Einstein--Hilbert action, which has led to Ho\v{r}ava--Lifshitz $F(\bar{R})$ \mbox{theories~\cite{Chaichian:2010yi,Elizalde:2010ep}}. These theories describe new scenarios that do not appear in standard $f(R)$ theories and thus they are relevant for further exploration. In the present article, we perform an analysis of the compatibility of these theories with the Swampland conjectures. In particular, we focus on the dS conjecture as explained before. 

The outline of this article is as follows. In Section \ref{S-SWFR}, we summarize the way in which standard $f(R)$ theories are studied in the light of the dS conjecture. We particularly focus on the way in which this conjecture can be made independent of $R$. In Section \ref{S-FRHL}, we present $F(\bar{R})$ theories, and in Section \ref{S-SWFRHL}, we perform the full analysis of their compatibility with the dS conjecture. Finally, in Section \ref{S-FinalRemarks}, we present our final remarks~section.
%%%%%%%%%%%%%%%%%%%%%%%%%%%%%%%%%%%%%%%%%%

\section{Swampland Conjectures in \boldmath{$f(R)$} Theories}
\label{S-SWFR}

In this section, we briefly overview the situation of the Swampland dS conjectures in the context of $f(R)$ theories.
Our aim is not to be exhaustive but just introduce conventions and notation which are necessary in the further sections. 

We consider the theory of gravity coupled to a scalar field. The dS conjecture was written originally in the following form (first dS conjecture) \cite{Obied:2018sgi,Ooguri:2018wrx,Andriot:2018wzk,Agrawal:2018own,Roupec:2018mbn}:
	\begin{equation}\label{dSConjecture}
		|V_{\phi}|\geq cV ,
	\end{equation}
where $V$ is the scalar field potential, $V_{\phi}$ is the derivative of the potential with respect to the scalar field which we are considering to be unique, and $c$ is an order $1$ constant. Later, this conjecture was refined to incorporate another possibility, which is written as (second \mbox{dS conjecture)}
	\begin{equation}\label{dSConjecture2}
		V_{\phi\phi}<-\widetilde{c}V ,
	\end{equation}
where $\widetilde{c}$ is another order $1$ constant.
Thus, the dS conjecture is expressed now as the statement that either one of the two inequalities, the first or the second dS conjecture, \mbox{must hold. }

Let us offer a general picture of the ways in which we can study the Swampland conjectures in the light of $f(R)$ theories \cite{Artymowski:2019vfy}. These theories are a generalization of GR by considering the gravitational action in the Jordan frame as\footnote{Throughout this work we will use units where $\kappa=8\pi G=1$}.
	\begin{equation}\label{ActionfR}
		S=\frac{1}{2}\int d^4x\sqrt{-g}f(R) ,
	\end{equation} 
where $f(R)$ is some well-behaved function of the curvature scalar $R$ and no matter component have been considered. %Check intended meaning retention.
 In this scenario, it is thought that the action for GR is just the first term on a Taylor expansion of the $f$ function around a small curvature. These theories are of great interest since they lead to many new interesting cosmological behaviours \cite{DeFelice:2010aj}. In order to transform this action into a form suitable to investigate its compatibility with the conjectures, we consider the equivalent action \cite{DeFelice:2010aj,Faraoni:2010pgm}
	\begin{equation}\label{EqActionfR}
		S=\frac{1}{2}\int d^4x\sqrt{-g}[\psi R-U],
	\end{equation}
where $\psi=\frac{df}{dR}=f_{R}$ is just an auxiliary field and $U=R\psi-f(R)$. Then, we can perform a conformal transformation in order to describe the dynamics in the Einstein frame in \mbox{the form} 
	\begin{equation}\label{ConfTransfR}
		\widetilde{g}_{\mu\nu}=e^{2\varphi}g_{\mu\nu} ,
	\end{equation}
with $\varphi=\frac{1}{2}\ln f_{R}$. We then obtain the following action:
	\begin{equation}\label{ActionfRF}
		S=\int d^4x\sqrt{-\widetilde{g}}\left[\frac{\widetilde{R}}{2}-\frac{1}{2}\widetilde{g}^{\mu\nu}\widetilde{\nabla}_{\mu}\phi\widetilde{\nabla}_{\nu}\phi-V(\phi)\right] ,
	\end{equation}
which represents just the action of gravity plus a scalar field where the scalar field is \mbox{given by}
	\begin{equation}\label{ScalarFieldfR}
		\phi=\sqrt{\frac{3}{2}}\ln f_{R} ,
	\end{equation}
with potential
	\begin{equation}\label{ScalarPotentialfR}
		V(\phi)=\frac{Rf_{R}-f}{2f^2_{R}} . 
	\end{equation}
Let us remark that in the Einstein frame, we have the action of gravity plus a canonically coupled scalar field, although in the Jordan frame, we did not consider any matter contribution. This scalar field originated in the process of reaching %Check intended meaning retention.
  the Einstein frame by a conformal transformation, and thus the field and its potential are entirely defined by geometry. However, since the dS conjecture is expected to hold for any scalar field that is canonically coupled to gravity, we can apply the conjecture in this scenario as well. We do not consider any matter content in the Jordan frame since, as we see by considering only gravity, we can use the conjectures to constraint the parameters of the gravitational theory and thus the allowed solutions.\footnote{This remark is similar to the scenario in string theory where the action in the string frame has not a canonically coupled scalar field and a conformal transformation is performed to the Einstein frame to obtain a canonically coupled scalar field. The only difference is that in our scenario the field is absent in the Jordan frame and it is defined by geometry.} We note from (\ref{dSConjecture}) that if the potential is negative, the conjecture is satisfied immediately. Therefore, we are interested in the scenario where the potential is positive. In this case, from (\ref{ScalarPotentialfR}), we see, therefore, that we have to impose the \mbox{condition}
\begin{equation}\label{ConditionfR}
	Rf_{R}-f>0 .
\end{equation}
And from (\ref{ScalarFieldfR}), we also note that we need $f_{R}>0$. Using these definitions, we can obtain the derivative of the potential with respect to the scalar field as
	\begin{equation}\label{fRDerivPotential}
		\frac{dV}{d\phi}=\frac{dV}{dR}\frac{dR}{df_{R}}\frac{df_{R}}{d\phi} .
	\end{equation}
 With this set up, the first dS conjecture (\ref{dSConjecture}) takes the general form 
\begin{equation}\label{dSConjfR}
	|2f-Rf_{R}|\geq\sqrt{\frac{3}{2}}c(Rf_{R}-f) ,
\end{equation}
and the second dS conjecture (\ref{dSConjecture2}) is written as
	\begin{equation}\label{dSConjfR2}
		\frac{f^2_{R}+f_{RR}(Rf_{R}-4f)}{3f_{RR}(Rf_{R}-f)}<-\widetilde{c} .
	\end{equation}
From these expressions, the conjectures were studied in a general way in \cite{Artymowski:2019vfy}, finding the constraints on the $f$ function that make it compatible with the conjectures. In \cite{Benetti:2019smr}, particular attention was put on a function of the form $f(R)=R^{1+\epsilon}$, where $\epsilon$ is a small number. It was found that the conjectures were compatible with the region of $\epsilon$ of phenomenological interest. Finally, in \cite{Elizalde:2022oej}, the first dS conjecture was found to be in good agreement with numerical data. 

We expect that in general  (\ref{dSConjfR}) will lead to inequalities involving $R$. However, since we want to use the conjectures to constraint the theory in this article, we are interested in a scenario where the conjecture could be made independent of $R$ and thus valid for all values of $R$. We can achieve this by imposing 
	\begin{equation}\label{CondionfRConstant}
		\frac{|2f-Rf_{R}|}{Rf_{R}-f}=A ,
	\end{equation}
where $A$ is any constant that should be positive in order to fulfil (\ref{ConditionfR}). Let us remark that this is an ansatz, and therefore it is not the only way to fulfil (\ref{dSConjfR}), but it is the only way to fulfil it independently of $R$. We  have therefore two possibilities regarding the sign of the term inside the absolute value. If $2f>Rf_{R}$, (\ref{CondionfRConstant}) leads to 
	\begin{equation}\label{fRConsForm}
		f(R)=\beta R^{\gamma} ,
	\end{equation}
where $\gamma=\frac{A+2}{A+1}$ with $1<\gamma<2$. In order to have a well-defined scalar field and not ghostly gravitons, we require that $f_{R}>0$, and in order to have positive mass of the curvature fluctuations, we also require $f_{RR}>0$ \cite{DeFelice:2010aj}. Expression (\ref{fRConsForm}) fulfills both conditions for $\beta>0$. Substituting this form in the first dS conjecture, we obtain that $\gamma$ is restricted to be
	\begin{equation}\label{fRRestrictGamma}
		\gamma\leq\frac{2+\sqrt{\frac{3}{2}}c}{1+\sqrt{\frac{3}{2}}c} .
	\end{equation}
In particular for $c=1$, this leads to $\gamma\lesssim1.45$. Thus, the conjecture leads to a restriction on the exponent of the $f$ function that cannot be the GR value but it is restricted to be close.

For the case $2f<Rf_{R}$, we obtain the same form, but in this case $\gamma=\frac{A-2}{A-1}$, and thus $\gamma>2$. This form can also fulfil the two conditions $f_{R}>0$ and $f_{RR}>0$ for $\beta>0$. However, in this range of values $\gamma$, the first dS conjecture cannot be fulfilled.

On the other hand, substituting the $f(R)$ function given by (\ref{fRConsForm}) into (\ref{dSConjecture2}), we determine that the second dS conjecture is written as
	\begin{equation}
		\frac{(\gamma-2)^2}{3(\gamma-1)^2}<-\widetilde{c} ,	
	\end{equation}
which is independent of $R$ but cannot be fulfilled for any value of $\gamma$. Therefore, this form for the $f$ function is incompatible with the second dS conjecture. This is not a problem since we only need to fulfill one of the two versions of the conjecture.

Thus, we conclude that in $f(R)$ theories, the only way to fulfill the first dS conjecture independently of $R$ is to take $f$ as a power term of $R$. The conjecture restricts the exponent of this function to be close to the GR value, and the second dS conjecture is never valid. 

In the following sections, we describe the $F(\bar{R})$ Ho\v{r}ava--Lifshitz theories and investigate whether a similar procedure as the one just described can be applied to them to look for the viability of the  Swampland conjectures.

\section{$F(\bar{R})$ Ho\v{r}ava--Lifshitz Theories}
\label{S-FRHL}

This section is devoted to a brief overview of the Ho\v{r}ava--Lifshitz theory of gravity in the context of modified gravity. We consider the ADM decomposition of spacetime in which we write the metric in the general form of
	\begin{equation}\label{MetricADM}
		ds^2=-N^2dt^2+g^{(3)}_{ij}(dx^{i}+N^{i}dt)(dx^{i}+N^{i}dt) ,
	\end{equation}
where $N$ is the lapse function, $N^{i}$ is the shift functions and $g^{(3)}_{ij}$ is the three metric. We note that the Friedmann--Lemaitre--Robertson-Walker (FLRW) metric has this form with vanishing shift functions, and it is always possible to choose the lapse function to be equal to one. 

In the Ho\v{r}ava--Lifshitz theory of gravity,  the metric is naturally written in an ADM form, and the action for the gravitational part can be written as  \cite{Horava:2009uw,Bertolami:2011ka,Sotiriou:2010wn,Kiritsis:2009sh,Mukohyama:2010xz}
	\begin{equation}\label{ActionHL}
		S_{HL}=\frac{1}{2}\int d^4xN\sqrt{g^{(3)}}N\left[K^{ij}K_{ij}-\lambda K^2+\mathcal{L}(g^{(3)}_{ij})\right] ,
	\end{equation}
where $K_{ij}$ is the extrinsic curvature, $\lambda$ is a parameter and $\mathcal{L}(g^{(3)}_{ij})$ is a potential term for gravity. In the general scenario, this term contains the scalar curvature of the three-metric $R^{(3)}$  and seven constants accompanying higher spatial derivative terms written in terms of the Ricci scalar of the spatial three metric. If the detailed balanced condition is proposed, this term is derived from a geometrical action and the number of constants is reduced. However, for the flat FLRW metric, which is of interest to us, this term is always zero since in that case the three metric is just the scale factor times the three-dimensional euclidean flat metric. Thus, in the following, we are not concerned with the specific form for this term. As can be easily seen, the IR limit is obtained in the limit $\lambda\to1$, and it is determined that % Check intended meaning retention
 $\mathcal{L}(g^{(3)}_{ij})\to R^{(3)}$. The lapse function appearing in the action can lead to different versions of the theory depending on the variables in which it may depend on. If the projectability condition is used, the lapse function only depends on the time variable, and thus the hamiltonian constraint is a global constraint, contrary to the case in which the lapse function is allowed to depend also on the spatial variables. Since we deal with the FLRW metric, we choose $N=1$, but in general, the global nature of the hamiltonian constraint leads to an integration constant, and thus a new term in the equations of motion absent in the other scenario \cite{Sotiriou:2010wn,Mukohyama:2009mz}. Thus, even though in both cases we can choose the lapse function to be one, the equations of motion will be different. This is an important remark for later on. 

In \cite{Wu:2019xtv}, the Swampland conjectures have been studied, when a nonlinear dispersion relation is taken into account. Such feature comes naturally from HL gravity and its possible extensions (such as \cite{Zhu:2011yu}). However, for an FLRW flat metric, the equations of motion are almost identical in all types of HL theories to the ones obtained in GR, and thus the problem of the incompatibility of the slow roll criteria is still present. Therefore, in the following, we consider a generalization of the HL theory in the same line as $f(R)$ theories and investigate the compatibility with the Swampland conjectures. 

As we can see from the form of action (\ref{ActionHL}), it is natural to generalize HL theories in the same way as the generalization leading to $f(R)$ theories described in the previous section. In this case, a more general action is proposed in the form
	\begin{equation}\label{ActionFBarR}
		S_{F(\bar{R})}=\int d^4x\sqrt{g^{(3)}}NF(\bar{R}) ,
	\end{equation}
where now $\bar{R}$ can be understood as a generalization of $R$ which includes the new terms of spatial derivatives which is proposed to have the form \cite{Chaichian:2010yi,Elizalde:2010ep}
	\begin{equation}\label{DefBarR}
		\bar{R}=K^{ij}K_{ij}-\lambda K^2+2\mu\nabla_{\rho}(n^{\rho}\nabla_{\nu}n^{\nu}-n^{\nu}\nabla_{\nu}n^{\rho})+\mathcal{L}(g^{(3)}_{ij}) ,
	\end{equation}
where $\mu$ is a constant. The term containing $\mu$ is usually omitted in the standard $f(R)$ proposal since it turns out to be a total derivative term. However, it is necessary for these theories. As we can see for the above definitions, the limit $\lambda\to1$, $\mu\to1$ of this theory leads to the standard $f(R)$ theory used in the previous section. This theory was originally proposed in a series of papers \cite{Chaichian:2010yi,Elizalde:2010ep} using the detailed balanced condition. The resulting cosmological scenarios were found to describe new and interesting behaviours for the flat FLRW metric such as solutions with two periods of accelerating expansion. In \cite{Lopez-Revelles:2012xal}, the Ekpyrotic scenario was studied for these theories. In \cite{Carloni:2010nx}, the detailed balanced condition was abandoned. However, for the flat FLRW metric, as we pointed out earlier, the equations of motion are the same. A hamiltonian analysis for these theories was performed in  \cite{Chaichian:2010zn}, and the relation to general scalar tensor HL theories was investigated in \cite{Kluson:2011ff}. An extension to include a $U(1)$ field was performed in \cite{Kluson:2010za}. However, as in the latter case, the equations of motion found for the FLRW flat metric were the same as in the original scenario with or without the detailed balanced condition. Thus, if we consider only the flat FLRW metric at the level of equations of motion, all the versions proposed for the $F(\bar{R})$ generalization will lead to the same results. Therefore, the following analysis is valid in all these scenarios.

Let us consider then a flat FLRW metric in the form
	\begin{equation}\label{FLRWFlatMetric}
		ds^2=-dt^2+a^2(t)\big[(dx^1)^2+(dx^2)^2+(dx^3)^2\big] .
	\end{equation}
With this metric, we obtain from (\ref{DefBarR}) the following result:
	\begin{equation}\label{BarRFLRW}
		\bar{R}=(3-9\lambda+18\mu)H^2+6\mu\frac{d}{dt}\left(H\right) ,
	\end{equation}
where $H=\frac{1}{a}\frac{da}{dt}$ is the Hubble parameter. In contrast, the standard curvature scalar in this case is
	\begin{equation}\label{RFLRW}
		R=12H^2+6\frac{d}{dt}\left(H\right) ,
	\end{equation}
and thus the difference between the two theories for this metric are just the appearance of the two parameters $\mu$ and $\lambda$, and therefore, at least for the flat FLRW metric, the interpretation of $\bar{R}$ can be considered as related to the curvature. However, we see that although both expressions differ only by constants, they lead to very important differences.

Considering only gravity, the first equation of motion obtained through the hamiltonian constraint is \cite{Chaichian:2010yi}
	\begin{equation}\label{EoM1}
		F(\bar{R})-6\left[(1-3\lambda+3\mu)H^2+\mu\dot{H}\right]F'(\bar{R})+6\mu H\frac{dF'(\bar{R})}{dt}-\frac{C}{a^3}=0 ,
	\end{equation}
where $\dot{H}=\frac{dH}{dt}$,  $F'=\frac{dF}{d\bar{R}}$ and where $C\neq0$ is an integration constant in the projectable version of the theory which  can be considered to represent dark matter for the way in which it appears in the equations of motion when it takes positive values \cite{Mukohyama:2009mz}. In the non-projectable version of the theory, we have $C=0$. This is the only difference between the two versions of the theory that is of importance to us. Varying with respect to the three metric, the second equation of motion is obtained: \cite{Chaichian:2010yi}	 
	\begin{equation}\label{EoM2}
		F(\bar{R})-2(1-3\lambda+3\mu)(\dot{H}+3H^2)F'(\bar{R})+2(3\lambda-1)H\frac{dF'(\bar{R})}{dt}+2\mu\frac{d^2F'(\bar{R})}{dt^2}=0 .
	\end{equation}
As stated before, we note from these equations and (\ref{DefBarR}) that the limit in which we can recover the standard $f(R)$ theories is $\mu,\lambda\to1$. If we put an ansatz of the form $H=H_{0}$ in the above equations and squared or cubic polynomial form for the $F(\bar{R})$ function, the cosmological solutions describe exponential acceleration for two different periods \cite{Chaichian:2010yi,Elizalde:2010ep}. This behaviour is exclusive of these theories since it is removed in the mentioned limit. This is one of the reasons that shows the importance of these kinds of theories.

Before continuing, let us remark on an important point. The Swampland conjectures were proposed originally in theories with Lorentz invariance. However, HL theories, as well as their generalizations, break this invariance in the UV, so one may worry about the applicability of the conjectures to these kinds of theories. If we apply the conjectures and find that they are incompatible, we may conclude that Lorentz invariance is a key requirement of the conjectures. However, in the next section, we discover that the conjectures are indeed compatible with these theories and they actually lead to some expected behaviours. Thus, we can support the idea that the conjectures may be applicable to more scenarios than originally thought considering our results as one indication a posteriori supporting this idea. In \cite{Trivedi:2021nss}, the compatibility of the conjectures with a cosmologcal setup with explicit breaking of Lorentz invariance was studied with positive results as well.

\section{Swampland Conjectures for \boldmath{$F(\bar{R})$} Theories}
\label{S-SWFRHL}

In the same way that we need to perform a conformal transformation to the $f(R)$ theory in order to obtain an action of GR plus  a scalar field in the Einstein frame, we perform a conformal transformation in the three metric of the above theories, and then it is possible to study the compatibility with the Swampland conjectures. Let us start by considering the action
	\begin{equation}\label{AuxActionBarF}
		S=\frac{1}{2}\int d^4x \sqrt{g^{(3)}}N\left[\bar{R}F'(\bar{R})-U\right] ,
	\end{equation}
with $U=\bar{R}F'(\bar{R})-F(\bar{R})$. This action is equivalent to (\ref{ActionFBarR}) with an auxiliary field. Then, we perform a conformal transformation only on the three metric of the form
	\begin{equation}\label{ConformalTrans3}
		g_{ij}^{(3)}=e^{-\bar{\phi}}\widetilde{g}^{(3)}_{ij} ,
	\end{equation}
and choose $N=1$ and $N^{i}=0$ as in the flat FLRW metric, which leads to the action in the Einstein frame \cite{Elizalde:2010ep,Carloni:2010nx} given by
	\begin{multline}\label{FbarRActionAux}
		S=\frac{1}{2}\int d^4x\sqrt{\widetilde{g}^{(3)}}\left[\widetilde{K}^{ij}\widetilde{K}_{ij}-\lambda \widetilde{K}^2+\left(-\frac{1}{2}+\frac{3\lambda}{2}-\frac{3\mu}{2}\right)\widetilde{g}^{(3)ij}\dot{\widetilde{g}}^{(3)}_{ij}\dot{\bar{\phi}}\right. \\ \left.+\left(\frac{3}{4}-\frac{9\lambda}{4}+\frac{9\mu}{2}\right)\dot{\bar{\phi}}^2 -\mathcal{L}(e^{-\bar{\phi}}g^{(3)}_{ij})-2V(\bar{\phi})\right],
	\end{multline}
with 
	\begin{equation}\label{AuxScalarField}
		\bar{\phi}=\frac{2}{3}\ln F'(\bar{R}) , \hspace{0.5cm} V(\bar{\phi})=\frac{\bar{R}F'-F}{2F'}. 
	\end{equation}
In order to eliminate the term that combines the metric with the scalar field and to obtain a canonically coupled scalar field (that is, that it appears as in standard scalar theories), \mbox{we choose}
	\begin{equation}\label{ElectioLambda}
		\mu=\lambda-\frac{1}{3}.
	\end{equation}
Thus, we perform the redefinition of the scalar field as
	\begin{equation}\label{RedScalarField}
		\phi=\alpha\bar{\phi}=\frac{2\alpha}{3}\ln F'(\bar{R}),
	\end{equation}
where 
	\begin{equation}\label{DefAlpha}
		\alpha=\sqrt{\frac{3}{4}-\frac{9\lambda}{4}+\frac{9\mu}{2}}=\frac{\sqrt{3(3\lambda-1)}}{2} .
	\end{equation}
Then, we finally obtain the action in the Einstein frame in the following form:
	\begin{equation}\label{ActionBarFRE}
		S=\int d^4x\sqrt{\widetilde{g}^{(3)}}\left[\frac{\widetilde{K}^{ij}\widetilde{K}_{ij}-\lambda \widetilde{K}^2-\mathcal{L}(e^{-\bar{\phi}}g^{(3)}_{ij})}{2}+\frac{\dot{\phi}^2}{2}-V(\phi)\right] .
	\end{equation}
Moreover, in the Einstein frame, we obtain an action of gravity plus a canonically coupled scalar field. We note that we do not obtain the HL theory of gravity since in this case there is no higher derivative terms of the scalar field; however, since the general form is the one of gravity plus a scalar field, we can use this action to investigate the compatibility with the Swampland conjectures of $F(\bar{R})$ theories. We do not worry that the gravity part of the action is not GR since we only need to describe a theory of gravity. For example, in~\cite{Yi:2018dhl}, the conjectures are studied with a Gauss--Bonnet term in the action. We note that since we made the choice (\ref{ElectioLambda}), we can no longer recover the $f(R)$ theories in any limit. We first study this case
since this simplification allows a complete analytic analysis. The general scenario where this special choice is not required is studied in Subsections \ref{SS-G} and \ref{SS-CH}.

With this setup and using (\ref{fRDerivPotential}) with $\bar{R}$, the first dS conjecture (\ref{dSConjecture}) is given in the general form
	\begin{equation}\label{dSConjectureBarFRAux}
		\frac{|V_{\phi}|}{V}=\frac{3}{2\alpha}\frac{|F|}{\bar{R}F'-F}\geq c ,
	\end{equation}
which can be written as
	\begin{equation}\label{dSConjectureBarFR}
		|F|\geq\frac{2\alpha c}{3}(\bar{R}F'-F) ,
	\end{equation}
and the condition for the scalar field to be positive is
	\begin{equation}\label{ConditionBarFR}
		\bar{R}F'-F>0.
	\end{equation}
On the other hand, the second dS conjecture takes the general form
	\begin{equation}\label{dSSecondConjectureBarFR}
		\frac{V_{\phi\phi}}{V}=\frac{9}{4\alpha^2}\frac{(F')^2-F''F}{F''(RF'-F)}<-\widetilde{c} .
	\end{equation}

We note that similarly to what happened with $f(R)$ theories, the way in which the above expressions are to be independent of $\bar{R}$ is by considering $F(\bar{R})$ as some power of $\bar{R}$. However, since these theories have not been studied as extensively as the $f(R)$ theories, we proceed in a different way. Instead of choosing some form for the $F$ function, we use the system of equations of motion in the Jordan frame with an ansatz for the Hubble parameter, which is of cosmological interest, and then construct the $F(\bar{R})$ function. Finally, we check whether this function can fulfill the conjectures at the same time that leads to cosmologically %Check intended meaning retention.
interesting solutions. 

Choosing (\ref{ElectioLambda}), the system of equations of motion (\ref{EoM1}) and (\ref{EoM2}) takes the form
	\begin{equation}\label{System1}
		F(\bar{R})+2(3\lambda-1)H\frac{dF'(\bar{R})}{dt}+\frac{2}{3}(3\lambda-1)\frac{d^2F'(\bar{R})}{dt^2}=0 ,
	\end{equation}
	\begin{equation}\label{System2}
		F(\bar{R})-2(3\lambda-1)\dot{H}F'(\bar{R})+2(3\lambda-1)H\frac{dF'(\bar{R})}{dt}-\frac{C}{a^3}=0 .
	\end{equation}
With this choice, we also obtain
	\begin{equation}\label{System3}
		\bar{R}=(3\lambda-1)(3H^2+2\dot{H}) .
	\end{equation}
From (\ref{System1}) and (\ref{System2}), we can obtain
	\begin{equation}\label{System4}
		\frac{d^2}{dt^2}F'(\bar{R})+3\dot{H}F'(\bar{R})+\frac{3C}{2(3\lambda-1)a^3}=0 .
	\end{equation}
Thus, the general procedure is to use this system of equations to construct the $F(\bar{R})$ function with an interesting cosmological behaviour. Since the constant $C$ changes the possible solutions, we consider  separately the two different scenarios, namely the case in which it can be eliminated, that is, the non-projectable case, and the case in which it has to be different from zero, which is the projectable case.

\subsection{Non-Projectable Case}
\label{SS-NP}
Let us  begin first with the non-projectable case in which $C=0$. We could start by proposing the simplest ansatz of cosmological interest, namely a constant Hubble parameter which describes a universe with an exponential acceleration ideal for an inflationary era. With this proposal in \cite{Chaichian:2010yi}, it was shown that these theories could result in a solution with two different periods of accelerated expansion. However, with the reduced system of equations (that is, after choosing the value of $\mu$), this proposal leads us to a vanishing value for the $F$ function. Therefore,  we propose instead an ansatz that describes an accelerated expanding universe but in the form of a power law with the time parameter; that is, we propose that the scale factor in the Jordan frame has the form
	\begin{equation}\label{ScaleFactorJordan}
		a(t)=t^n ,
	\end{equation}
where $n$ is a constant that is considered as positive. Thus, the Hubble parameter is $H=n/t$. With this ansatz using Equation (\ref{System3}), we determine that $\bar{R}$ is related to the time variable as
	\begin{equation}\label{BRTime}
		\bar{R}=\frac{(3\lambda-1)n(3n-2)}{t^2} .
	\end{equation}
We note from (\ref{RedScalarField}) that we must always have $3\lambda>1$ in order to have the scalar field properly defined. We are generally interested in values of $n$ that describe accelerating expanding universes and therefore, from the above, we note that considering positive values for $\bar{R}$, we obtain the condition $3n-2>0$. We note from this expression that $\bar{R}$ grows inversely with time with a behaviour similar to what we could expect of the curvature.  In this scenario, Equation (\ref{System4}) is written as
	\begin{equation}\label{NPFEq}
		t^2\frac{d^2}{dt^2}F'(\bar{R})-3nF'(\bar{R})=0 ,
	\end{equation}
which leads to solution
	\begin{equation}\label{NPFAux}
		F'(\bar{R})=c_{1}t^{\alpha_{+}}+c_{2}t^{\alpha_{-}} ,
	\end{equation}
where $c_{1}$ and $c_{2}$ are integration constants and
	\begin{equation}\label{NPAlphas}
		\alpha_{\pm}=\frac{1}{2}\left[1\pm\sqrt{1+12n}\right] .
	\end{equation}
The other equations on the system are immediately satisfied. Using (\ref{BRTime}), we can find the form of $F$ as a function of $\bar{R}$. In this way, we obtain
	\begin{equation}\label{NPFBR}
		F(\bar{R})=-\frac{A_{1}}{\beta_{+}\bar{R}^{\beta_{+}}}+\frac{A_{2}}{\beta_{-}}\bar{R}^{\beta_{-}} ,
	\end{equation}
where we defined the positive constants
	\begin{equation}\label{NPBetas}
		\beta_{+}=\frac{\alpha_{+}}{2}-1=\frac{1}{4}\left[\sqrt{1+12n}-3\right] \geq0 , \hspace{0.5cm} \beta_{-}=1-\frac{\alpha_{-}}{2}=\frac{1}{4}\left[\sqrt{1+12n}+3\right]>\frac{3}{2}.
	\end{equation}
The last inequality follows from the condition $3n-2>0$ in both cases, and we also defined 
\begin{equation}
	A_{1}=c_{1}\left[(3\lambda-1)n(3n-2)\right]^{\beta_{+}+1} , \hspace{0.5cm} A_{2}=\frac{c_{2}}{\left[(3\lambda-1)n(3n-2)\right]^{\beta_{-}-1}} .
\end{equation}

We note that our general solution contains two terms of powers of $\bar{R}$, and thus we expect that it can fulfill the dS conjecture for any of the terms taken independently. The condition to have a non-negative potential for the scalar field in the Einstein frame (\ref{ConditionBarFR}) in this case takes the form
	\begin{equation}\label{NPCondition}
		\bar{R}F'-F=\frac{A_{1}}{\bar{R}^{\beta_{+}}}\left(1+\frac{1}{\beta_{+}}\right)+A_{2}\bar{R}^{\beta_{-}}\left(1-\frac{1}{\beta_{-}}\right)>0.
	\end{equation}
Furthermore, from (\ref{NPBetas}), we obtain that 
	\begin{equation}\label{BFBetasP}
		1+\frac{1}{\beta_{+}}=\frac{\sqrt{1+12n}+1}{\sqrt{1+12n}-3}>0 , \hspace{0.5cm} 1-\frac{1}{\beta_{-}}=\frac{\sqrt{1+12n}-1}{\sqrt{1+12n}+3}>0  .
	\end{equation}
Thus, condition (\ref{NPCondition}) can be easily fulfilled by taking positive values for the integration constants $c_{1}$ and $c_{2}$. In this case, we obtain
	\begin{equation}\label{NPAbsF}
		|F|=\bigg\rvert-\frac{A_{1}}{\beta_{+}\bar{R}^{\beta_{+}}}+\frac{A_{2}}{\beta_{-}}\bar{R}^{\beta_{-}}\bigg\rvert .
	\end{equation}
Since there is a minus sign in the first term, the $F$ function cannot have a definite sign for all values of $\bar{R}$, and thus we cannot fulfill the dS conjecture for all values of $\bar{R}$ if we consider both terms at the same time as we anticipated. Thus, let us consider each term separately. 

First, let us consider the positive power factor on $F(\bar{R})$; that is, we choose $c_{1}=0$ and $c_{2}>0$, and then the first dS conjecture (\ref{dSConjectureBarFR}) leads to
	\begin{equation}\label{dSConjecturNPPositive}
		\frac{1}{3}<\lambda\leq\frac{1}{3}+\frac{16}{c^2(\sqrt{1+12n}-1)^2} .
	\end{equation}
In the other case, if we consider the negative power factor on $F(\bar{R})$ by choosing $c_{1}>0$ and $c_{2}=0$, the conjecture leads to
	\begin{equation}\label{dSConjecturNPNegative}
		\frac{1}{3}<\lambda\leq\frac{1}{3}+\frac{16}{c^2(\sqrt{1+12n}+1)^2}.
	\end{equation}

In both cases condition (\ref{ConditionBarFR}) is satisfied. Thus, the dS conjecture leads, in both cases, to an inequality for the HL parameter $\lambda$. We also note that the faster we want the expansion to be, that is, the bigger the value of $n$, the closer we are moved to the value $\lambda=1/3$. Thus, in order to fulfill the conjecture independently of $\bar{R}$ and to have a fast expansion, we determine that $\lambda$ must be bigger but close to $1/3$ and thus away for its IR limit value. We also note that since, in both cases, the first dS conjecture leads to a region of validity for the $\lambda$ parameter and neither $\beta_{+}$ nor $\beta_{-}$ in (\ref{NPFBR}) depend on $\lambda$, the form of the $F(\bar{R})$ function is not constrained by the conjecture, it only depends on $n$. Thus, we have the freedom to choose any positive values of interest for these terms, in contrast to the standard $f(R)$ case.

On the other hand, the second dS conjecture (\ref{dSSecondConjectureBarFR}) in this case is written in general as
	\begin{equation}\label{FBRdS2General}
	\begin{array}{l}
		\dfrac{V_{\phi\phi}}{V}\vspace{3pt}\\
		=\dfrac{9}{4\alpha^{2}}\dfrac{\left[\dfrac{A_{1}}{\bar{R}^{\beta_{+}+1}}+A_{2}\bar{R}^{\beta_{-}-1}\right]^2-\left[-\dfrac{A_{1}}{\beta_{+}\bar{R}^{\beta_{+}}}+\dfrac{A_{2}}{\beta_{-}}\bar{R}^{\beta_{-}}\right]\left[-\dfrac{A_{1}(\beta_{+}+1)}{\bar{R}^{\beta_{+}+2}}+A_{2}(\beta_{-}-1)\bar{R}^{\beta_{-}-2}\right]}{\left[-\dfrac{A_{1}(\beta_{+}+1)}{\bar{R}^{\beta_{+}+2}}+A_{2}(\beta_{-}-1)\bar{R}^{\beta_{-}-2}\right]\left[\dfrac{A_{1}}{\bar{R}^{\beta_{+}}}\left(1+\dfrac{1}{\beta_{+}}\right)+A_{2}\bar{R}^{\beta_{-}}\left(1-\frac{1}{\beta_{-}}\right)\right]} \vspace{3pt}\\<-\widetilde{c}.
		\end{array}
	\end{equation}
Taking the positive powers of $\bar{R}$, that is, taking $c_{1}=0$, we obtain 
	\begin{equation}\label{DBRdS2P}
		\frac{9}{4\alpha^{2}}\frac{1}{(\beta_{-}-1)^2}<-\widetilde{c} ,
	\end{equation}
which can never be fulfilled because the term on the left hand side is always positive. Taking now the negative powers of $\bar{R}$, that is, taking $c_{2}=0$, we obtain condition
		\begin{equation}\label{DBRdS2N}
		\frac{9}{4\alpha^{2}}\frac{1}{(\beta_{+}+1)^2}<-\widetilde{c} ,
	\end{equation}
which, again, is never fulfilled. Thus, we determine that the second dS conjecture is never compatible with the obtained $F(\bar{R})$ function.

In summary, we determined that in these theories, we can describe universes with a power law scale factor that describes expanding universes in the Jordan frame, and this behaviour leads, naturally, to power law terms for the $F(\bar{R})$ function. In the Einstein frame, we obtain a scalar field in which we can apply the Swampland conjectures. The first one of the dS conjectures can be fulfilled for each term on the solution of $F$ independently of $\bar{R}$, leading to an inequality for the HL parameter $\lambda$ which is in agreement with the difficulties of achieving this scenario in GR since it leads us to the opposite of the infrared limit;  that is, it leads us to UV limit $\lambda\to1/3$ for rapidly expanding universes. We also determine that the second dS conjecture can never be valid for this form of the $F(\bar{R})$ function.

Finally, we explore what is the behaviour described by these solutions on the Einstein frame where we have the scalar field coupled to gravity. For this, we note that since both frames are related by the transformation (\ref{ConformalTrans3}) in the Einstein frame, we also have a flat FLRW metric, but now the scale factor is given by
	\begin{equation}\label{NPScaleFactorEF}
		\widetilde{a}(t)=t^n\left[c_{1}t^{\frac{1}{2}\left(1+\sqrt{1+12n}\right)}+c_{2}t^{\frac{1}{2}\left(1-\sqrt{1+12n}\right)}\right]^{1/3}.
	\end{equation}
Let us consider each term separately, as performed previously. For the negative powers of $\bar{R}$, we choose $c_{1}=1$ and $c_{2}=0$. Thus, the scale factor is 
\begin{equation}\label{ScaleFactorEFN}
	\widetilde{a}(t)=t^{\frac{1}{6}\left[6n+1+\sqrt{1+12n}\right]}=t^{\theta_{+}} .
\end{equation}
We note that $\theta_{+}$ does not depend on $\lambda$, and therefore result (\ref{dSConjecturNPNegative}) derived from the dS conjecture does not restrict the range of values that it can take. We determine, in this case, that the exponent is always positive, and thus the universe is always expanding. 

\begin{figure}[h!]
	 	\includegraphics[width=0.8\textwidth]{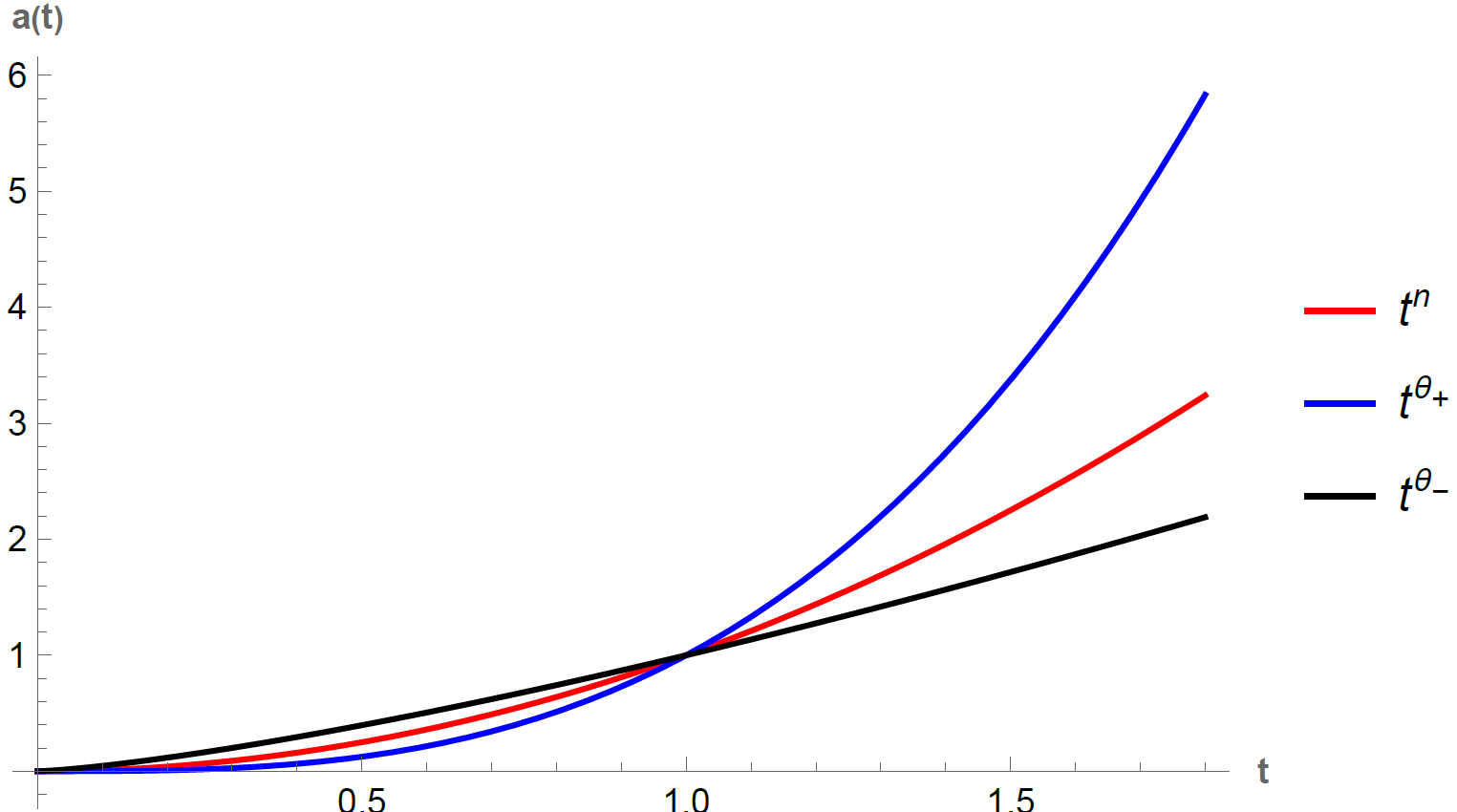}
	 	\caption{Scale factor with $n=2$ in the Jordan frame (Red curve) and in the Einstein frame with negative powers of $\bar{R}$ in $F$ (Blue curve) and with positive powers (Black curve).}
	 	\label{ScaleFactorPC}
	 \end{figure}

On the other hand, for the positive power term of $\bar{R}$, we choose $c_{1}=0$ and $c_{2}=1$. Thus, the scale factor is given by
	\begin{equation}\label{ScaleFactorEFP}
		\widetilde{a}(t)=t^{\frac{1}{6}\left[6n+1-\sqrt{1+12n}\right]}=t^{\theta_{-}} .
	\end{equation}
Once again, $\theta_{-}$ is independent of $\lambda$ and always takes positive values describing expanding universes. It can be shown that in general,
	\begin{equation}\label{DiffThetas}
		\theta_{-}<n<\theta_{+} .
	\end{equation}
Thus, for negative powers of $\bar{R}$ in $F$, we have, in the Einstein frame, a more rapidly expanding universe than in the Jordan frame for $t>1$. On the contrary, for positive powers, we have a slower-expanding universe. For small time values $t<1$, this behaviour is flipped, and the fastest expanding universe corresponds to positive powers of $\bar{R}$. We show the behaviour just described  for $n=2$ in Figure  \ref{ScaleFactorPC}.

%%%%%%%%%%%%%%%%%%%%%%%%%%%%% 
%%%%%%%%%%%%%%%%%%%%%%%%%%%%%
\subsection{Projectable Case}
\label{SS-P}
Let us consider now the projectable case in which we take $C\neq0$ in the system of Equations (\ref{System1})--(\ref{System4}). We propose the same ansatz as before, that is, the scale factor as a power law on the time variable. Thus, Equation (\ref{BRTime}) is still valid. In this case, Equation (\ref{System4}) leads to
	\begin{equation}\label{EqAuxP}
		\frac{d^2}{dt^2}F'(\bar{R})-\frac{3n}{t^2}F'(\bar{R})+\frac{3C}{2(3\lambda-1)t^{3n}}=0 .
	\end{equation}
The solution of this equation consists of the homogeneous solution (the same as before with $C=0$) and a particular solution. In this form, we obtain solution
	\begin{equation}\label{BFTimeP}
		F'(\bar{R})=c_{1}t^{\alpha_{+}}+c_{2}t^{\alpha_{-}}-\frac{3C}{2(3\lambda-1)(9n^2-12n+2)}t^{2-3n},
	\end{equation}
where we note that we must exclude the possible roots of the polynomial $9n^2-12+2$, which are $n\approx0.1952,1.138$. In this case, the function $F(\bar{R})$ takes the form
	\begin{equation}\label{BarRFN}
		F(\bar{R})=-\frac{A_{1}}{\beta_{+}\bar{R}^{\beta_{+}}}+\frac{A_{2}}{\beta_{-}}\bar{R}^{\beta_{-}}-A_{C}\bar{R}^{3n/2} ,
	\end{equation}
where we defined
	\begin{equation}
		A_{C}=\frac{C}{(3\lambda-1)n(9n^2-12n+2)\left[2(3n-2)(3\lambda-1)\right]^{3n/2-1}}.
	\end{equation}

The condition for the positivity of the scalar field potential (\ref{ConditionBarFR}) leads, in this case, to
	\begin{equation}\label{PotentialPositiveProyectable}
		\bar{R}F'-F=\frac{A_{1}}{\bar{R}^{\beta_{+}}}\left(1+\frac{1}{\beta_{+}}\right)+A_{2}\bar{R}^{\beta_{-}}\left(1-\frac{1}{\beta_{-}}\right)-A_{C}\bar{R}^{3n/2}\left(\frac{3n-2}{2}\right)>0 .
	\end{equation}
Since Equation (\ref{BFBetasP}) holds, we note that in order to fulfill this condition for all values of $\bar{R}$, we need $A_{C}<0$; thus, $\frac{C}{9n^2-12n+2}<0$. Therefore, we have two possibilities:
	\begin{itemize}
		\item $C>0$ and $\frac{2}{3}<n<\frac{2+\sqrt{2}}{3}$.
		\item $C<0$ and $n>\frac{2+\sqrt{2}}{3}$ .
	\end{itemize}
We note that in this case,
	\begin{equation}\label{PAbsF}
		F(\bar{R})=\bigg\rvert-\frac{A_{1}}{\beta_{+}\bar{R}^{\beta_{+}}}+\frac{A_{2}}{\beta_{-}}\bar{R}^{\beta_{-}}-A_{C}\bar{R}^{3n/2}\bigg\rvert .
	\end{equation}
Since the last term will always be present, if we take $c_{2}=0$,  function $F$ will not have a definite sign for all values of $\bar{R}$ and thus will not fulfill the dS conjecture for all values of $\bar{R}$. Therefore, we consider only the positive power and take $c_{1}=0$. In this case, we can fulfill the dS conjecture for all positive values of $\bar{R}$ but not independently of it as we stated before. The first dS conjecture, in this case, leads to
	\begin{equation}\label{dSConjectureProyectable}
		A_{2}\bar{R}^{\beta_{-}}\left[\frac{1}{\beta_{-}}-\frac{2\alpha c}{3}\left(1-\frac{1}{\beta_{-}}\right)\right]-A_{C}\bar{R}^{3n/2}\left[1-\frac{2\alpha c(3n-2)}{6}\right]\geq0 .
	\end{equation}
Moreover, we note that in order to fulfill this inequality for all values of $\bar{R}$, we must ask each term within square brackets to be positive. In this form, we obtain the two inequalities
	\begin{equation}\label{PLambda1}
		\lambda\leq\frac{1}{3}+\frac{16}{c^2(\sqrt{1+12n}-1)^2} ,
	\end{equation}
	\begin{equation}\label{PLambda2}
		\lambda\leq\frac{1}{3}+\frac{4}{c^2(3n-2)^2} .
	\end{equation} 
We note that (\ref{PLambda1}) is the same as the one obtained in the non-projectable case (\ref{dSConjecturNPPositive}). However, since both inequalities are on the same parameter, we only need to impose the stronger one. It turns out that (\ref{PLambda1}) is more restrictive than (\ref{PLambda2}) only if $n<\frac{2+\sqrt{2}}{3}\approx1.138$. Thus, for most of the values of $n$, the dS conjecture is satisfied by (\ref{PLambda2}) for every value of $\bar{R}$. Then, considering $C\neq0$, we are lead to a stronger condition for parameter $\lambda$ for most cases. %Check intended meaning retention.

From (\ref{BarRFN}) and taking $c_{1}=0$, we note that in this case,
	\begin{equation}
		F'=A_{2}\bar{R}^{\beta_{-}-1}-\frac{3n}{2}A_{C}\bar{R}^{\frac{3n}{2}-1} ,
	\end{equation}
	\begin{equation}
		F''=A_{2}(\beta_{-}-1)\bar{R}^{\beta_{-}-2}-\frac{3n}{4}(3n-2)A_{C}\bar{R}^{\frac{3n}{2}-2} .
	\end{equation}
Thus, since $-A_{C}>0$, we have $F''>0$ for all $\bar{R}>0$. However, from these expressions, we have
	\begin{multline}
		(F')^2-F''F=\frac{A_{2}^2}{\beta_{-}}\bar{R}^{2\beta_{-}-2}+\frac{3n}{2}A^{2}_{C}\bar{R}^{3n-2}\\+A_{2}A_{C}\left[-3n+\frac{3n}{2\beta_{-}}\left(\frac{3n-2}{2}\right)+(\beta_{-}-1)\right]\bar{R}^{3n/2+\beta_{-}-2},
	\end{multline}
which does not have a definite sign for all values of $\bar{R}$. Therefore, the second dS conjecture (\ref{dSSecondConjectureBarFR}) cannot be studied in general for all values of $\bar{R}$ in this case.

In summary, in the projectable case, the first dS conjecture can still be fulfilled for all positive values of $\bar{R}$, but not independently of it. The conjecture also leads to an inequality for the $\lambda$ parameter, and for most of the values of $n$, this condition is more restrictive than in the non-projectable case. 

\begin{figure}[h!]
		\includegraphics[width=0.8\textwidth]{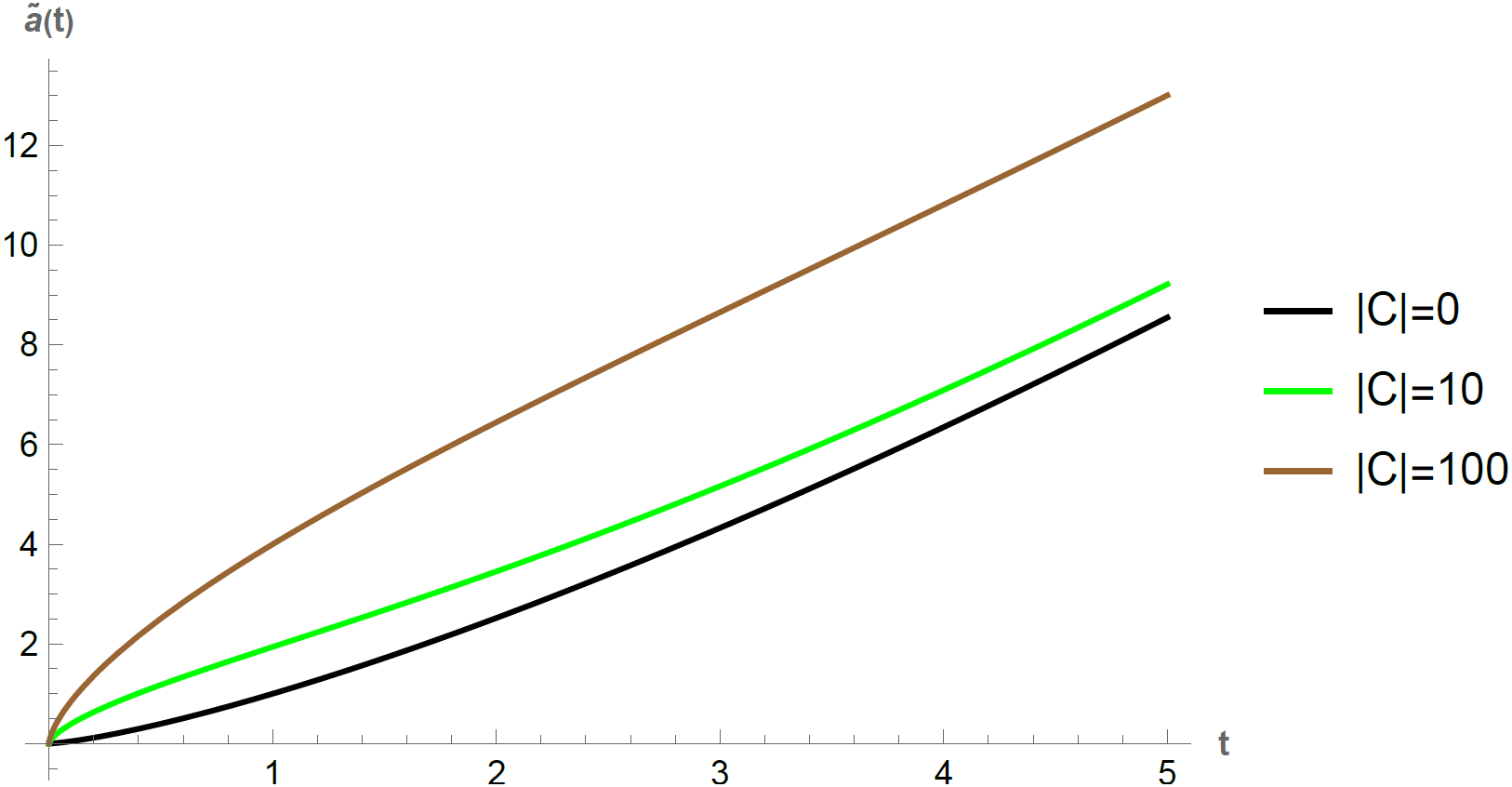}
		\caption{Scale factor in the Einstein frame for the projectable case with $\lambda=0.39$, $c=1$, $n=2$ and choosing $C=-|C|$ for $|C|=0$ (Black curve) (Non-projectable case), $|C|=10$ (Green curve) and $|C|=100$ (Brown curve). In all cases, we have expanding universes.}
		\label{ScaleFactorPr}
	\end{figure}

Once again, the metric in the Einstein frame is also a flat FLRW metric. In this case, taking $c_{2}=1$, the scale factor is 
	\begin{equation}\label{PScaleFactorEins}
		\widetilde{a}=t^n\left[t^{\frac{1}{2}(1-\sqrt{1+12n})}-\frac{3C}{2(3\lambda-1)(9n^2-12n+2)}t^{2-3n}\right]^{\frac{1}{3}} .
	\end{equation}
Thus, in the projectable case, the scale factor has a dependence on $\lambda$ and the dS conjecture leads to a lower bound for it. Since both terms are positive, the universe is always expanding. In Figure \ref{ScaleFactorPr}, we plot this scale factor choosing $\lambda=0.39$, $c=1$ and $n=2$ for different values of $C$. Since $n=2$, in order to fulfill the conditions mentioned, we need to take negative values for $C$; thus, we write $C=-|C|$. We see that the effect of projectability, that is, of values for $|C|$ different from zero, is that as $|C|$ increases, the scale factor increases, making the expansion faster.

Finally, let us note from (\ref{BarRFN}) that taking $c_{1}=0$ and $c_{2}=1$, we have two terms of positive powers of $\bar{R}$. Thus, we can hope to write this function in a form that resembles an approximation of an Einstein--Hilbert term plus corrections. We also note from the dS conjecture inequality (\ref{PLambda1}) that if $n$ is close to its limiting value $2/3$, we can have access to the IR limit $\lambda\to1$. Thus, let us study the form that our solutions take for $n$ close to $2/3$. We write
	\begin{equation}\label{Approxn}
		n=\frac{2}{3}(1+\epsilon) ,
	\end{equation}
with $\epsilon<<1$. In this case, we can take $C>0$, and the most restrictive inequality is  (\ref{PLambda1}). From (\ref{BarRFN}), taking $c_{1}=0$, $c_{2}=1$ and (\ref{Approxn}), we obtain
	\begin{multline}\label{BFREpsilon}
		F(\bar{R})=\frac{3C}{4(1+\epsilon)|1-2\epsilon^2|(4\epsilon)^{\epsilon}(3\lambda-1)^{\epsilon+1}}\bar{R}^{\epsilon+1}\\+\frac{4}{[\sqrt{9+8\epsilon}+3]\left[\frac{4}{3}\epsilon(\epsilon+1)(3\lambda-1)\right]^{(\sqrt{9+8\epsilon}-1)/4}}\bar{R}^{(\sqrt{9+8\epsilon}+3)/4} .
	\end{multline}
For $\epsilon<<1$, we can make the approximation \cite{Benetti:2019smr}
	\begin{equation}\label{Approx}
		\bar{R}^{1+\epsilon}=\sum_{\nu=0}^{\infty}\frac{\epsilon^{\nu}\bar{R}\ln^{\nu}(\bar{R})}{\nu!}\simeq\bar{R}+\epsilon\bar{R}\ln(\bar{R}) .
	\end{equation}
Thus, we obtain 
	\begin{equation}\label{BFREpsilonF}
		F(\bar{R})\approx\frac{3C}{4(1+\epsilon)|1-2\epsilon^2|(4\epsilon)^{\epsilon}(3\lambda-1)^{\epsilon+1}}\bar{R}+F_{c}(\bar{R}),
	\end{equation}
with the correction term
	\begin{multline}\label{BFRCorrec}
		F_{c}(\bar{R})=\frac{3C\epsilon}{4(1+\epsilon)|1-2\epsilon^2|(4\epsilon)^{\epsilon}(3\lambda-1)^{\epsilon+1}}\bar{R}\ln\bar{R}\\ +\frac{4}{[\sqrt{9+8\epsilon}+3]\left[\frac{4}{3}\epsilon(\epsilon+1)(3\lambda-1)\right]^{(\sqrt{9+8\epsilon}-1)/4}}\bar{R}^{(\sqrt{9+8\epsilon}+3)/4} .
	\end{multline}
Consequently, the scale factor in the Einstein frame (\ref{PScaleFactorEins}) takes the form
	\begin{equation}
		\bar{a}(t)=t^{\frac{2}{3}(1+\epsilon)}\left[\frac{1}{t^{\frac{1}{2}(\sqrt{9+8\epsilon}-1)}}+\frac{3C}{4(3\lambda-1)|1-2\epsilon^2|}\frac{1}{t^{2\epsilon}}\right]^{\frac{1}{3}} .
	\end{equation}
In this case, we can take values of $\lambda$ greater than one and the resulting universes are always expanding. However, let us point out a subtle issue with this IR limit. From (\ref{System3}), we obtain
	\begin{equation}
		\lim_{\lambda\to1}\bar{R}=6H^2+4\dot{H} .
	\end{equation}
We note that this value does not coincide with the GR value of $R$ (\ref{RFLRW}). Thus, since we chose $\mu=\lambda-\frac{1}{3}$, the $\lambda\to1$ IR limit does not correspond to GR. In order to perform a correct GR limit in Section \ref{SS-G}, we avoid this choice and consider the conjectures in the general case.

\subsection{Constraints on $\lambda$}
\label{SS-Lamdas}
\begin{figure}[h!]
	\includegraphics[width=0.9\textwidth]{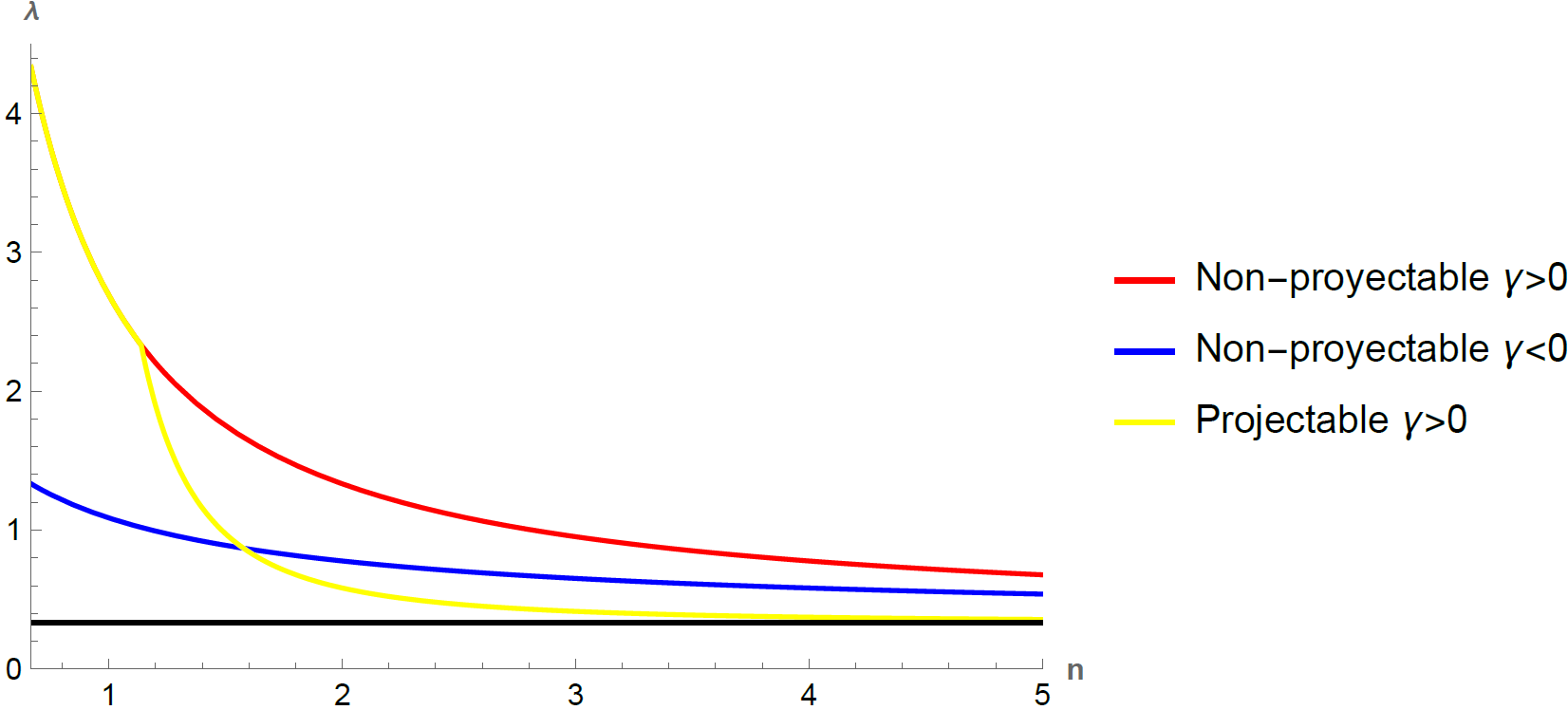}
	\caption{Allowed values for $\lambda$ in agreement with the dS conjecture with $c=1$. The black horizontal line indicates the lower bound $1/3$. For the non-projectable case, we have $F(\bar{R})\sim\bar{R}^{\gamma}$. The upper bounds for the non-projectable case are shown when $\gamma<0$ (Blue curve) and when $\gamma>0$ (Red curve). For the projectable case, we have the additional term in $F(\bar{R})$ with $C\neq0$ and $\gamma>0$. The corresponding upper bound is shown as well (Yellow curve). }
	\label{Lambdas}
\end{figure}

As we showed in the previous subsections, in the particular case of $\mu=\lambda-1/3$, the first dS conjecture leads  to constraints on the $\lambda$ parameter in both the projectable and the non-projectable cases.
In order to visualize in a graphical form these constraints, we plot, in Figure \ref{Lambdas}, for the non-projectable case, the upper values  (\ref{dSConjecturNPPositive}) and (\ref{dSConjecturNPNegative}) for different values of $n$. We note that values of $\lambda$ bigger than one are allowed only when $n$ is small, but as $n$ grows, the allowed region is reduced in both cases. In general, we note that for positive power terms in the $F(\bar{R})$ function, the allowed region is bigger than for  negative power terms. For the projectable case, we show the behaviour of the corresponding upper bound coming from (\ref{PLambda1}) and (\ref{PLambda2}). We choose the most restrictive constraint as we vary $n$. It is shown that for small values of $n$, the constraint is the same as the one obtained in the non-projectable case. However, when $n$ is large enough, the constraint coming from this scenario is more restrictive, and thus the allowed range of $\lambda$ becomes smaller and decreases inversely with $n$. In all cases, we choose $c=1$.

\subsection{The General Case}
\label{SS-G}
In the general scenario in which we consider $\mu$ and $\lambda$ in  (\ref{FbarRActionAux}) as independent parameters and the scalar field is defined by (\ref{RedScalarField}). We note that in general, we do not have an action that can be divided into a sum of an action for gravity plus the action of a scalar field because of the third term in the action that combines the metric with the scalar field. However, we do have an explicit definition for a scalar field with the correct form of the kinetic term in the action as before, and we also have a definition for its corresponding potential. Thus, we propose that the dS conjectures are still applicable for this action. A similar scenario was presented in \cite{Brahma:2019kch}, were the conjectures were studied with a scalar field that is non-canonically coupled to gravity as well.

%%%%%%%%%%%%%%%%%%%%%%%%%%%%%%%%%%%%%%%%%%%%%%%%%%%
%%%%%%%%%%%%%%%%%%%%%%%%%%%%%%%%%%%%%%%%%%%%%%%%%%%
%%%%%%%%%%%%%%%%%%%%%%%%%%%%%%%%%%%%%%%%%%%%%%%%%%%

In order to have a correct definition of the scalar field, we need parameter $\alpha$ defined in (\ref{DefAlpha}) to be a real number which leads to
	\begin{equation}\label{D1}
		\lambda<\frac{1}{3}+2\mu.
	\end{equation}
From the system of Equations (\ref{EoM1}) and (\ref{EoM2}), we obtain
	\begin{equation}\label{GEoMEx}
		\frac{d^2F'}{dt^2}+\frac{3\lambda-1-3\mu}{\mu}H\frac{dF'}{dt}+\frac{3\lambda-1}{\mu}\dot{H}F'+\frac{C}{2\mu a^3}=0 .
	\end{equation}
We consider $\mu>0$ and $\lambda>0$, we also use the same ansatz for the scale factor in the Jordan frame as before, that is, $a(t)=t^n$, and consider the non-projectable case ($C=0$). Thus, the latter reduces to
	\begin{equation}\label{GEoM}
		t^2\frac{d^2F'}{dt^2}+\frac{3\lambda-1-3\mu}{\mu}nt\frac{dF'}{dt}-\frac{(3\lambda-1)n}{\mu}F'=0 .
	\end{equation}

With this ansatz from (\ref{BarRFLRW}), we obtain 
	\begin{equation}\label{GRbart}
		\bar{R}=\frac{D}{t^2} ,
	\end{equation}
with $D=3n[n(6\mu+1-3\lambda)-2\mu]$. As a result, we can still interpret $\bar{R}$ as related to the curvature. In the particular case, we considered only positive values for $\bar{R}$ which led us to a lower bound for $n$ compatible with expanding universes. For consistency and later convenience, we also ask, in the general case, for positive values of $\bar{R}$. Consequently, we obtain condition $D>0$ which leads to
	\begin{equation}\label{D2}
		\lambda<\frac{1}{3}+\frac{2\mu(3n-1)}{3n}  ,
	\end{equation}
from which we obtain $n>1/3$. Thus, as a result of considering only a correct definition of the scalar field and positive values for $\bar{R}$, we obtain upper bounds for $\lambda$. Now, let us proceed as in the particular case by solving (\ref{GEoM}) and construct the $F(\bar{R})$ function. The general solution of (\ref{GEoM}) is
	\begin{equation}\label{GF'}
		F'(\bar{R})=c_{1}t^{\delta_{+}}+c_{2}t^{\delta_{-}} ,
	\end{equation}
with
	\begin{equation}\label{DefDeltas}
		\delta_{\pm}=\frac{1}{2}\left[3n+1-B\pm\sqrt{(B+1-3n)^2+12n}\right] ,
	\end{equation}
where we define
\begin{equation}\label{DefB}
	B=\frac{3\lambda-1}{\mu}n .
\end{equation}
Using (\ref{GRbart}), we can integrate the latter expression to obtain
	\begin{equation}\label{GF}
		F(\bar{R})=-\frac{c_{1}D^{\delta_{+}/2}}{\gamma_{+}\bar{R}^{\gamma_{+}}}+\frac{c_{2}D^{\delta_{-}/2}}{\gamma_{-}}\bar{R}^{\gamma_{-}} ,
	\end{equation}
with $c_{1,2}$ constants of integration and 
	\begin{equation}\label{DefGammas}
		\gamma_{+}= -1 + \frac{\delta_{+}}{2} , \hspace{0.5cm} \gamma_{-}=1-\frac{\delta_{-}}{2}  .
	\end{equation}
The condition for the scalar field potential to be positive in this case takes the form
	\begin{equation}\label{GCondtionPotential}
		\bar{R}F'-F=\frac{c_{1}D^{\delta_{+}/2}}{\bar{R}^{\gamma_{+}}}\left(1+\frac{1}{\gamma_{+}}\right)+c_{2}D^{\delta_{-}/2}\bar{R}^{\gamma_{-}}\left(1-\frac{1}{\gamma_{-}}\right)>0 ,
	\end{equation}
which is the same as in (\ref{NPCondition}) with $\beta_{\pm}$ substituted by $\gamma_{\pm}$. Therefore, in order to fulfill this condition for all values of $\bar{R}$, we ask each term to be positive. We choose $c_{1}>0$ and $c_{2}>0$, and we are left with conditions $\delta_{+}>2$ and $\delta_{-}<0$. The first condition is fulfilled with (\ref{D2}), whereas the second leads to 
	\begin{equation}\label{D3}
		\lambda>\frac{1}{3} .
	\end{equation}
From (\ref{D2}) and (\ref{D3}), it can be proven that we always have $\gamma_{\pm}>0$, and therefore the $F(\bar{R})$ function in the general case (\ref{GF}) always consists of a negative power term of $\bar{R}$ and one with positive exponent. The negative power term has a negative coefficient, whereas the positive %Check intended meaning retention.
 power term has a positive coefficient as in the particular case studied before.

Thus, in the general case, we have three inequalities (\ref{D1}), (\ref{D2}) and (\ref{D3}) that constrain the values of $\lambda$. Since the three are constraints on the same parameter, it suffices to take the most restrictive one. Then, these inequalities lead to
	\begin{equation}\label{D3C}
		\frac{1}{3}<\lambda<\frac{1}{3}+\frac{2\mu(3n-1)}{3n} .
	\end{equation}

We note that the closer $\mu$ is to zero, the closer $\lambda$ becomes to its limiting value $1/3$. We also note that $\mu\neq0$ in order to correctly fulfill the inequalities, and therefore the simpler versions of $F(\bar{R})$ theories that do not take into account this term are inconsistent with this conformal transformation. Thus, the generalization of \cite{Chaichian:2010yi} is needed. %Check intended meaning retention.

The first dS conjecture has the form (\ref{dSConjectureBarFR}) with the $F(\bar{R})$ function (\ref{GF}). In order to fulfill this conjecture independently of $\bar{R}$, we take, once again, each term separately. For the negative power term, we take $c_{2}=0$, and the conjecture leads to
	\begin{equation}\label{D4}
		\sqrt{\frac{3\mu(6n-B)}{n}}\left[\sqrt{(B+1-3n)^2+12n}+3n+1-B\right]\leq\frac{12}{c} .
	\end{equation}
On the other hand, for the positive power term, we choose $c_{1}=0$, and the conjecture leads~to
	\begin{equation}\label{D5}
		\sqrt{\frac{3\mu(6n-B)}{n}}\left[\sqrt{(B+1-3n)^2+12n}+B-3n-1\right]\leq\frac{12}{c} .
	\end{equation}
In general, both expressions (\ref{D4}) and (\ref{D5}) lead to a region of validity for the $\lambda$ parameter in terms of $\mu$, and thus we can compare the resulting bounds with (\ref{D3C}) to investigate the region of compatibility. However, the expressions are complicated to solve analytically, and thus we carry out a numerical analysis.

\begin{figure}[h!]
	\includegraphics[width=0.8\textwidth]{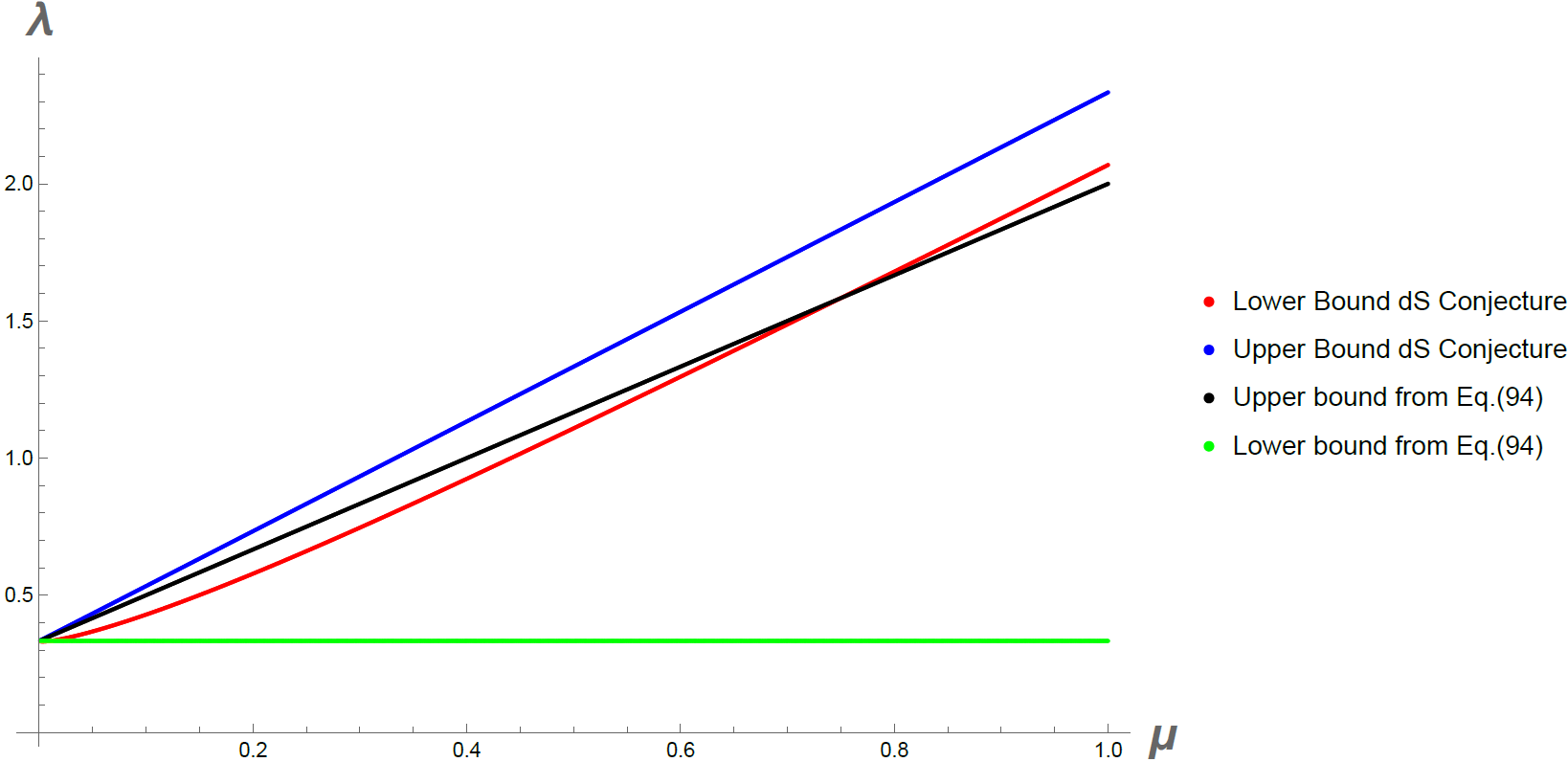}
	\caption{Allowed values for $\lambda$ in the non-projectable general scenario for the negative power term in $F(\bar{R})$ for $n=2$ and $c=2$. The lower and upper bounds from the dS conjecture are shown with the red and blue curves, respectively. The lower ($\lambda=1/3$) and upper bounds coming from Equation (\ref{D3C}) are also shown with the green and black curves, respectively. The allowed values of $\mu$ are restricted to the region when the red curve shows smaller values than the black one.}
	\label{GeneralLambdasN}
\end{figure}

For the negative power term, upper and lower bounds coming from (\ref{D4}) are found numerically. In Figure \ref{GeneralLambdasN}, these bounds are shown for $n=2$, $c=2$ and the varying $\mu$ parameter. %Check intended meaning retention.
 We also show the upper and lower bounds coming from (\ref{D3C}). We note that in order to fulfill both expressions, we need the upper bound from (\ref{D3C}) to be bigger than the lower bound from (\ref{D4}), and thus $\mu$  is bounded from above. Consequently, the allowed values of $\lambda$ increase as $\mu$ increases, but not above a maximum value. For smaller values of $n$ or bigger values of $c$, the restriction on $\mu$ is more severe, and $\lambda$ becomes closer to $1/3$.

\begin{figure}[h!]
\includegraphics[width=0.8\textwidth]{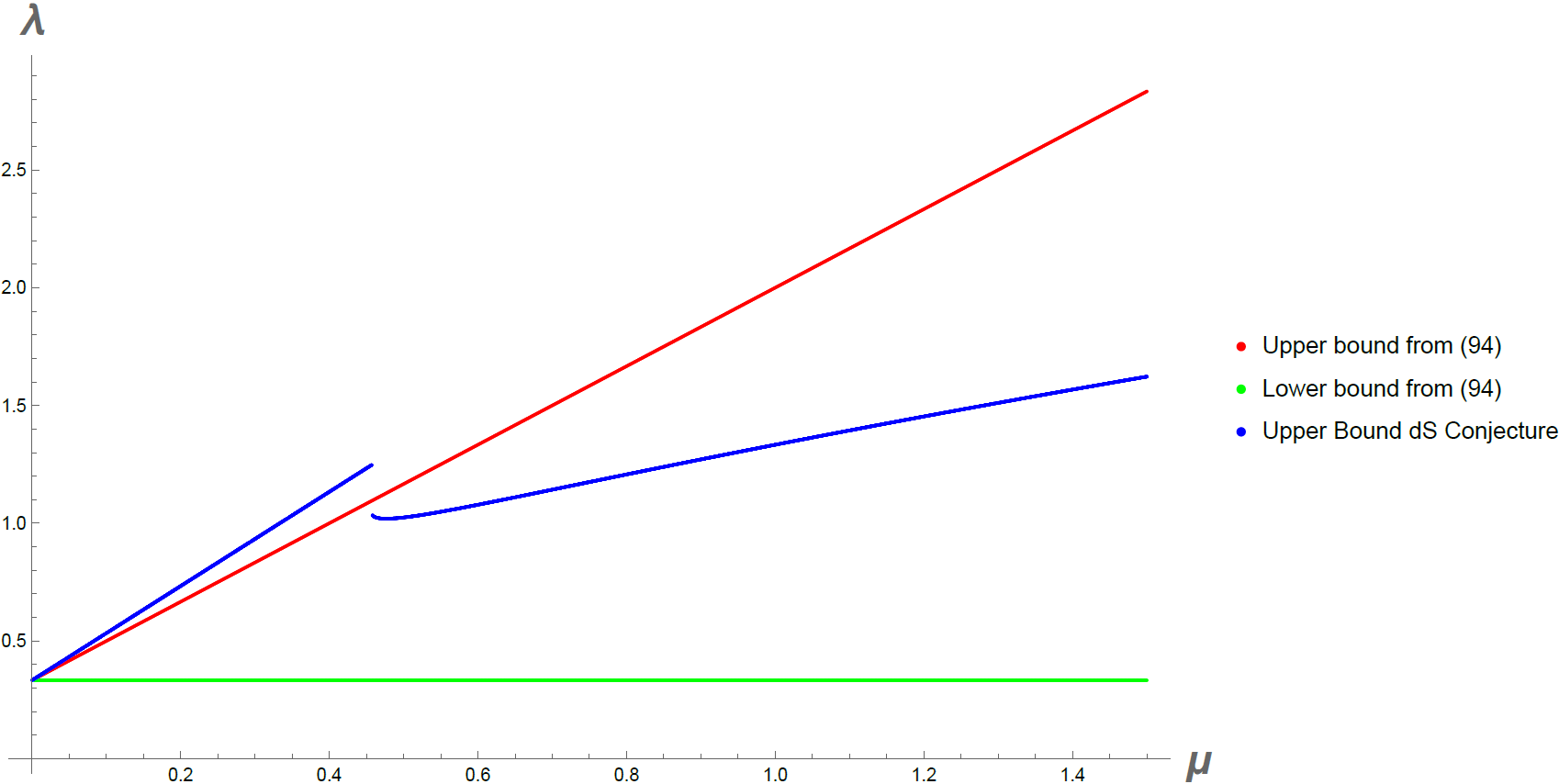}
\caption{Allowed values for $\lambda$ in the non-projectable general scenario for the positive power term in $F(\bar{R})$ for $n=2$ and $c=1$. The lower ($\lambda=1/3$) and upper bounds from (\ref{D3C}) are shown with the green and red curves, respectively. The upper bound from the dS conjecture is shown with the blue curve. For large enough values of $\mu$, there are two allowed regions. We show only the lowest upper bound since the other region is inconsistent with (\ref{D3C}).}
		\label{GeneralLambdasP}
	\end{figure}

For the positive power term in $F(\bar{R})$, we find, with numerical analysis, that small values of $\mu$ (\ref{D5}) lead to an upper bound for $\lambda$ that is bigger than the one coming from~(\ref{D3C}), and thus the conjecture is satisfied as a consequence of (\ref{D3C}). On the other hand, for values of $\mu$ that are large enough, (\ref{D5}) leads to two regions of validity, one is an upper bound which is smaller than (\ref{D3C}) and one is a region with an upper and lower bounds bigger than (\ref{D3C}) and therefore inconsistent. Thus, for large values of $\mu$, the dS conjecture is more restrictive than (\ref{D3C}). In any case, both upper bounds grow with $\mu$ and thus, in this scenario, we can have access to larger values of $\lambda$. It only becomes restricted to be close to $1/3$ if $\mu$ is small enough. In Figure \ref{GeneralLambdasP}, we show this behaviour for $n=2$ and $c=1$. For smaller values of $n$ or larger values of $c$, the region where the dS conjecture is more restrictive than (\ref{D3C}) is found for smaller values of $\mu$.

The second dS conjecture leads to (\ref{FBRdS2General}) with $\gamma_{\pm}$ instead of $\beta_{\pm}$ and thus, for the positive power term, we have (\ref{DBRdS2P}) with $\gamma_{-}$, and for the negative power term, we have (\ref{DBRdS2N}) with $\gamma_{+}$. Therefore, the second dS conjecture is never fulfilled in the general scenario either.

In the general case, the metric in the Einstein frame also has the FLRW form with the scale factor given by
	\begin{equation}\label{GSFAux}
		\widetilde{a}(t)=t^n\left[c_{1}t^{\delta_{+}}+c_{2}t^{\delta_{-}}\right]^{1/3} .
	\end{equation}
Thus, for the negative power term, we choose $c_{1}=1$ and $c_{2}=0$, and the scale factor takes the form 
	\begin{equation}\label{GSFN}
		\widetilde{a}(t)=t^{\theta_{+}} ,
	\end{equation}
with
	\begin{equation}\label{DefTheta+}
		\theta_{+}=\frac{1}{6}\left[9n+1-B+\sqrt{(B+1-3n)^2+12n}\right] .
	\end{equation}
Similarly, for the positive power term, we choose $c_{1}=0$ and $c_{2}=1$, and the scale factor takes the form
	\begin{equation}\label{GSFP}
		\widetilde{a}(t)=t^{\theta_{-}} ,
	\end{equation}
with
	\begin{equation}\label{DefTheta-}
		\theta_{-}=\frac{1}{6}\left[9n+1-B-\sqrt{(B+1-3n)^2+12n}\right] .
	\end{equation}
In the general scenario, the scale factor depends on $\lambda$ and $\mu$, and therefore the inequality (\ref{D3C}) and the region coming from the first dS conjecture restrict it. Moreover, it can be proven that we always have $\theta_{\pm}>0$, and thus we always have expanding universes.

Thus, the general scenario is similar to the particular case; that is, in order to have a properly defined scalar field and to fulfill the first dS conjecture, $\lambda$ becomes restricted in terms of $\mu$ around $1/3$, and the resulting description in the Einstein frame is expanding universes. For the positive power term, however, we can have access to larger values of $\lambda$ by increasing the values of $\mu$; this is important to consider the limit $\lambda,\mu\to1$ of interest in the following. %Check intended meaning retention. 
In this case, we also determine that the second dS conjecture is never fulfilled. 

Furthermore, the general scenario allows us correct performance of the limit $\mu,\lambda\to1$ which, as stated before, should correspond to standard $f(R)$. In this limit, (\ref{D1}) and (\ref{D3}) are automatically satisfied, whereas (\ref{D2}) leads to $n>1/2$. The fist dS conjecture corresponds to upper bounds for $n$. 

For the negative power term, (\ref{D4}) leads to $c\lesssim0.87$, which is inconsistent with $c$ being an order $1$ constant. For the positive power term, (\ref{D5}) leads to
	\begin{equation}\label{GAVn}
		\frac{1}{2}<n\leq\frac{\sqrt{3}}{c}\left(\frac{c+\sqrt{3}}{2c-\sqrt{3}}\right) , \hspace{0.5cm}\text{with}\hspace{0.5cm} 1\leq c\lesssim3.46 .
	\end{equation}
The largest upper value is obtained in $c=1$ and corresponds to approximately $17.66$. Using (\ref{DefDeltas}), we determine the allowed values of the exponent in the $F$ function to be %Check intended meaning rtention.
	\begin{equation}\label{GAVdelta}
		\gamma_{-}\big\rvert_{n=1/2}=1.25<\gamma_{-}\leq\gamma_{-}\big\rvert_{n=17.66}\simeq1.866 .
	\end{equation}
The allowed region (\ref{GAVdelta}) must be compared to the one obtained in the standard $f(R)$ case~(\ref{fRRestrictGamma}). The lower bound from (\ref{GAVdelta}) comes from the dependence of $\delta_{-}$ on $n$ and the condition $n>1/2$ coming from (\ref{D2}). Since the system of Equations (\ref{EoM1}) and (\ref{EoM2}) reduces to standard $f(R)$ in the limit $\mu,\lambda\to1$, $C\to0$, after proposing the ansatz $a(t)=t^n$, the same form of the $f$ function (\ref{GF}) should emerge, and thus, after imposing $D>0$ in the standard $f(R)$ theories, we obtain the same lower bound. The upper bound, in contrast, is different. This comes from the fact that although in the Jordan frame the $F(\bar{R})$ theory recovers the standard $f(R)$ in the mentioned limit, the transformation to the Einstein frame for $F(\bar{R})$ theories is only on the three metric (\ref{ConformalTrans3}), whereas in the standard $f(R)$, the conformal transformation used is in the complete four metric (\ref{ConfTransfR}). Thus, the resulting scalar field and scalar potential are different, and therefore the first dS conjecture has a different form. It is interesting, however, that even though the analytic forms of both upper bounds are different, their maximum values are not too far. In standard $f(R)$, we obtain approximately $1.45$, whereas in $F(\bar{R})$, we obtain approximately $1.866$. Thus, the $F(\bar{R})$ theory allows a bigger region for the exponent, although it is still not too far from one. Therefore, in the limit $\mu,\lambda\to1$, the $F(\bar{R})$ theories lead to a consistent result with the standard $f(R)$ theories, that is, the power of $\bar{R}\to R$ becomes restricted to be bigger than one (actually bigger than $1.25$), but not too big, smaller than approximately $1.866$. From (\ref{DefTheta-}), we determine that in this limit, the scale factor has the form of (\ref{GSFP})  with 
	\begin{equation}
		\frac{1}{3}<\theta_{-}\lesssim 17.083 .
	\end{equation}
Thus, there is enough room for accelerated expansion in the Einstein frame.

\begin{figure}[h!]
		\includegraphics[width=0.8\textwidth]{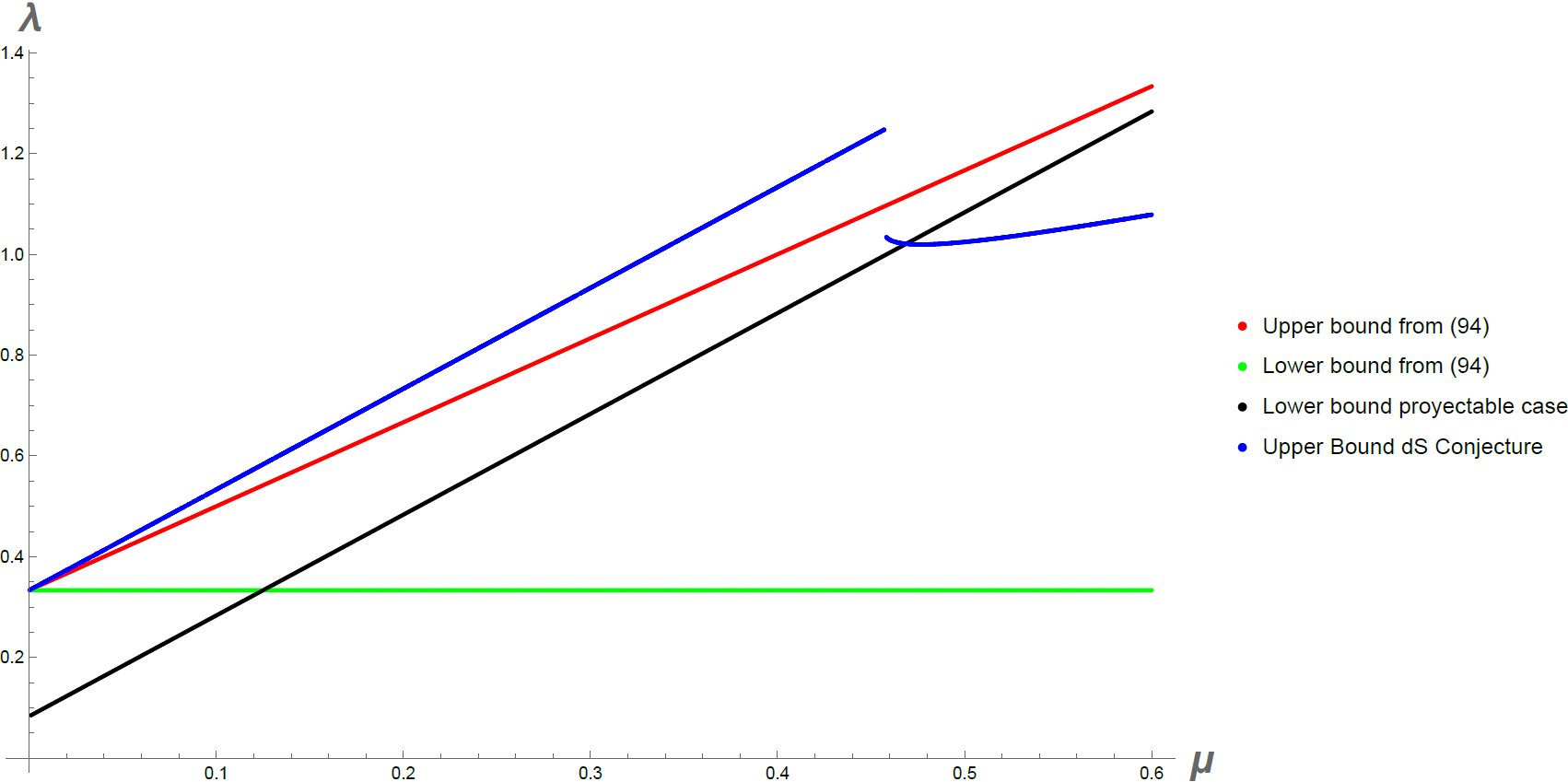}
		\caption{Allowed values for $\lambda$ in the general scenario with proyectability for $n=2$ and $c=1$. The lower ($\lambda=1/3$) and upper bounds from (\ref{D3C}) are shown with the green and red curves, respectively. The upper bound from the dS conjecture is shown with the blue curve. The lower bound from (\ref{DProy}) is shown with the black curve. The allowed values for $\mu$ are restricted to the region when the black curve displays smaller values than the blue or red curves.}
		\label{GeneralLambdasPPr}
	\end{figure}

Finally, let us study the projectable case. Since the obtained form for the $F$ function has the same form as the one obtained in the particular case, it only changes the specific form of the exponents. The projectable case can be treated as was performed in Section \ref{SS-P}. In this case, the $F$ function is written as
	\begin{equation}\label{FProjectableGeneral}
			F(\bar{R})=-\frac{c_{1}D^{\delta_{+}/2}}{\gamma_{+}\bar{R}^{\gamma_{+}}}+\frac{c_{2}D^{\delta_{-}/2}}{\gamma_{-}}\bar{R}^{\gamma_{-}}-\frac{C}{3\mu n(18n^2-15n+B-3nB+2)D^{3n/2-1}}\bar{R}^{3n/2} .
	\end{equation}
The condition for a positive potential is written analogously to (\ref{PotentialPositiveProyectable}), and it can also be fulfilled for all values of $\bar{R}$ by taking the coefficient for the new term to be positive and $n>2/3$. Once again, we take $c_{1}=0$; then, the first dS conjecture is written with two terms as in (\ref{dSConjectureProyectable}), and thus, in order to fulfill the conjecture for all values of $\bar{R}$, we obtain two inequalities. The first is the same as in (\ref{D5}), whereas the second is
	\begin{equation}\label{DProy}
		\lambda\geq\frac{1}{3}+2\mu-\frac{4}{(3n-2)^2c^2} .
	\end{equation}
Thus, we obtain an extra inequality that has to be taken at the same time with (\ref{D3C}) and (\ref{D5}), but now it is a lower bound for $\lambda$. In Figure \ref{GeneralLambdasPPr}, we show the bounds shown earlier for (\ref{D3C}) and (\ref{D5}), but we now add the bound coming from (\ref{DProy}) for $n=2$ and $c=1$. We note that in order to fulfill the complete system of inequalities, there is an upper bound for $\mu$, and thus $\lambda$ becomes restricted once again to be close to $1/3$. 

In the $f(R)$ limit, (\ref{DProy}) reduces to
	\begin{equation}
		n\leq\frac{1}{\sqrt{3}c}+\frac{2}{3} .
	\end{equation}
The upper bound acquires its maximum value at $c=1$ and corresponds to approximately $1.244$. Thus, in the proyectable case, the upper bound on $n$ is more restrictive than in the non-proyectable case, which leads to
	\begin{equation}\label{GRGProyectable}
		\gamma_{-}\big\rvert_{n=2/3}=1.295<\gamma_{-}\leq\gamma_{-}\big\rvert_{n=1.244}\simeq1.407 .
	\end{equation}
Thus, we obtain bounds that are even more restrictive than standard $f(R)$ in this case. However, the form of the $F$ function has an extra power term with an exponent of $3n/2$. If we take small values of $n$ around $2/3$, we can write this function as GR plus correction terms as in~(\ref{BFREpsilonF}) (the only difference is the coefficients of each term and the exponent of the second term in (\ref{BFRCorrec})), since in this case, we can obtain, consistently, $\bar{R}\to R$ in the mentioned~limit. 

\subsection{Constant Hubble Parameter}
\label{SS-CH}
Let us remark that in the particular case considered in Sections \ref{SS-NP}--\ref{SS-Lamdas}, it was not possible to obtain a solution with a constant Hubble parameter, and thus we used the ansatz~(\ref{ScaleFactorJordan}). In the general case, we keep this ansatz as in the particular case, where  we found constraints on the parameters of the theory. However, in the general case, there is not a restriction of this sort since, as we know from reference \cite{Chaichian:2010yi}, this kind of solution does exist when we do not relate the $\mu$ and $\lambda$ parameters. In this subsection, we investigate this scenario on the light of the dS conjecture as well. 

If the Hubble parameter is constant, we note from (\ref{BarRFLRW}) that
\begin{equation}\label{ConstantBarR}
	\bar{R}=3(1-3\lambda+6\mu)H^2 ,
\end{equation}
and thus it is also a constant; therefore, $F(\bar{R})$ does not depend on time. Thus, both Equations~(\ref{EoM1}) and (\ref{EoM2}) lead to 
\begin{equation}\label{ConstantEoM}
	F(\bar{R})-6H^2(1-3\lambda+3\mu)F'(\bar{R})=0 ,
\end{equation}
where we set $C=0$ for consistency of the equation. Since $\bar{R}$ is a constant, this equation does not lead to a unique solution. In \cite{Chaichian:2010yi}, some forms for the $F(\bar{R})$ function were proposed, and two periods of accelerating expansion were obtained. However, we note from (\ref{dSConjectureBarFR}) that if we want to fulfill the dS conjecture without a dependence on $\bar{R}$ as well as use this to constraint the theory, as it was performed on the previous sections, we should select the form $F(\bar{R})=A\bar{R}^{\gamma}$ with $A$ and $\gamma$ constants. Therefore, with this ansatz, we obtain from (\ref{ConstantBarR}) \mbox{and (\ref{ConstantEoM})} 
\begin{equation}
	\gamma=\frac{3\lambda-1-6\mu}{2\left(3\lambda-1-3\mu\right)} .
\end{equation}
It can be shown that we can fulfill the positivity of the potential (\ref{ConditionBarFR}) and obtain a real value for $\alpha$ (\ref{DefAlpha}) with the condition
\begin{equation}\label{ConstantCondition1}
	\lambda<\frac{1}{3}+\mu ,
\end{equation}
and thus $\bar{R}>0$ always. The first dS conjecture  takes the form of (\ref{dSConjectureBarFR}), and, in this case, leads~to
\begin{equation}\label{ConstantdSConjecture}
	\frac{(3\lambda-1)\sqrt{1-3\lambda+6\mu}}{2\sqrt{3}(3\mu-3\lambda+1)}\leq\frac{1}{c} .
\end{equation}
Let us remark for the  standard $f(R)$ limit $\mu,\lambda\to1$ that (\ref{ConstantCondition1}) is satisfied, but (\ref{ConstantdSConjecture}) leads to $c<\frac{\sqrt{3}}{2}\simeq0.866$, and thus it is not compatible with $c$ being an order 1 constant. In general, there are two regions where (\ref{ConstantCondition1}) and (\ref{ConstantdSConjecture}) can be satisfied given by
\begin{equation}
	\lambda<\frac{1}{3} , \hspace{1cm} \mu\geq\frac{3\lambda-1}{6} ,
\end{equation}
or
\begin{equation}
	\lambda>\frac{1}{3} , \hspace{1cm} \mu\geq-\frac{1}{3}+\frac{c^2(3\lambda-1)^2}{36}+\lambda+\frac{1}{36}\sqrt{c^2(3\lambda-1)^3\left[12+c^2(3\lambda-1)\right]} .
\end{equation} 
Thus, we can fulfill the dS conjecture if $\lambda<1/3$ and $\mu$ takes any positive value or $\lambda>1/3$ and $\mu$ is bounded from below with a value that grows with $\lambda$ and $c$. In particular, we cannot take the limit $\mu,\lambda\to1$ in any of the two regions. Furthermore, we note from (\ref{ConformalTrans3}) that the scale factor in the Einstein frame takes the form
\begin{equation}
	\tilde{a}(t)=\left[\frac{3\lambda-1-6\mu}{2(2\lambda-1-3\mu)}\left[\left(3-9\lambda+18\mu\right)H^2\right]^{\frac{3\lambda-1}{2(3\mu-3\lambda+1)}}\right]e^{Ht} ,
\end{equation}
and thus we have an exponentially expanding universe in the Einstein frame as well. Therefore, a constant Hubble parameter is compatible with the dS conjecture and leads to restrictions over the parameters of the theory that forbids the infrared behaviour $\lambda,\mu\to1,$ making this compatibility exclusive to Ho\v{r}ava--Lifshitz $F(\bar{R})$ theories.

\section{Final Remarks}
\label{S-FinalRemarks}
In the present article, we presented an analysis of the compatibility of the Ho\v{r}ava--Lifshitz $F(\bar{R})$ theories with the Swampland conjectures, focusing on the dS conjecture and using a flat FLRW metric.

We first presented the analysis for the standard $f(R)$ theories which was previously performed. In the Jordan frame, we had the $f(R)$ action and we needed to perform a conformal transformation to the Einstein frame where an action of gravity was obtained plus a canonically coupled scalar field. It was in this frame where the Swampland conjectures could be applied to the scalar field, which is defined in terms of the $f(R)$ function. We particularly focused on the way in which the dS conjecture can be made independent of $R$. It was found that the only way in which such a scenario can be obtained is when the $f$ function is a power term on $R$. With this form, the dS conjecture and consistency of the theory led to conditions on the power of $R$ that has to be greater than $1$ (consequently, not GR) but close to it since the upper bound was found to be around $1.45$. In this scenario, the second dS conjecture could never be fulfilled. 

We then moved on to study the Ho\v{r}ava--Lifshitz $F(\bar{R})$ theories that represent a generalization to the HL gravity theory in the same way as standard $f(R)$ theories generalize GR. The analysis was performed using a flat FLRW metric, and thus it is valid for all the different versions of this theory, and its generalizations whose equations of motion for this metric are the same. In order to obtain a scalar field in the Einstein frame, a conformal transformation was also performed, but this time only on the three metric as it has been previously established. In this sense, a scalar field can be obtained, and its corresponding potential in terms of the $F(\bar{R})$ function. From this, it was easily seen that in order to fulfill the conjecture independently of $\bar{R}$, we would also need a power term in the $F$ function. However, the general strategy was to propose an ansatz for the scale factor in the form of $a(t)=t^n$, consider positive values of $\bar{R}$, and then use the equations of motion to construct the $F$ function and finally investigate the compatibility with the dS conjecture. However, in the Einstein frame, the action contained an extra term that was written in terms of derivatives of the three metric and of the scalar field. This term could be eliminated with an appropriate choice of the parameters of the theory, putting $\mu$ in terms of $\lambda$ and thus staying with only one parameter left. 

We first studied this particular case that allowed us an action of gravity plus a scalar field canonically coupled. Thus, we were able to apply the dS conjecture. The projectability condition of the $F(\bar{R})$ theory was found to be relevant, since it adds an extra term in the equations of motion. We first studied the non-projectable case in which this term can be ignored. In this case, the $F(\bar{R})$ function was found to consist of two power terms, one with a positive exponent and positive coefficient and one with a negative exponent and negative coefficient. In order to fulfill the dS conjecture independently of $\bar{R}$, we studied each term separately. In both cases, the conjecture led to upper bounds for the $\lambda$ parameter around its UV value of $1/3$. The faster the expansion, the closer $\lambda$ reached this extreme value. In this case, the second dS conjecture was also never fulfilled. In the projectable case, there was an extra positive power term in the $F$ function, and thus the conjecture could not be fulfilled independently of $\bar{R}$. However, it could be fulfilled for all positive values of $\bar{R}$. The conjecture led to another upper bound for $\lambda$, which made it closer to $1/3$. We showed these upper bounds for the two cases in Figure \ref{Lambdas}. In this case, the second dS conjecture could not lead to a result for all values of $\bar{R}$. In both cases, the metric in the Einstein frame had the same flat FLRW form, and it was always expanding. In the non-projectable case, the scale factor only depended on $n$, and we showed the general behaviour in Figure \ref{ScaleFactorPC}. It was found that for $t<1$, the most rapidly expanding universe is the one with the positive power term and the least rapidly expanding one is the one with the negative power; however, for $t>1$, the situation was reversed. For the projectable case, the scale factor was dependent on $\lambda$, and thus it was subjected to the dS conjecture. In Figure \ref{ScaleFactorPr} we showed this behaviour. It was found that the scale factor increases as $|C|$ increases.

Finally, we proposed that in the general scenario in which $\mu$ and $\lambda$ were considered as independent parameters and the action could not be put in the form of gravity plus a scalar field, the dS conjecture can still be applicable since we do have a well-defined scalar field and its corresponding potential. By performing the same procedure as before, we considered the non-projectable case and found the same general form for the $F$ function, but the exponents were found to depend on $\mu$ and $\lambda$. In this case, there were more inequalities to consider in order to have a correct description in the Einstein frame, which led to upper bounds for the $\lambda$ parameter in terms of the $\mu$ parameter. If $\mu\to0$, these bounds produced $\lambda\to1/3$. The dS conjecture could be studied in general for each term on the $F$ function and led to regions of validity in terms of $\lambda$ and $\mu$, but in order to find regions of compatibility of both expressions, we had to rely on a numerical analysis. In Figure \ref{GeneralLambdasN}, we showed these bounds for different values of $\mu$ for the negative power term. It was found that in order to fulfill both inequalities, an upper bound for $\mu$ appeared, and thus $\lambda$ became restricted around $1/3$ as well. These upper bounds were found to become more restrictive as $n$ decreased or $c$ increased. However, in Figure  \ref{GeneralLambdasP},  we showed these bounds for different values of $\mu$ for the positive power term. In this case, the bounds of $\lambda$ could become bigger as $\mu$ increased. Due to this behaviour, it was possible to perform the $f(R)$ limit $\mu,\lambda\to1$. In this limit, it was found that the only possible form of $F(\bar{R})$ compatible with the dS conjecture was a positive exponent which has to be greater than  $1.25$ and lesser than around $1.866$. The lower bound came from the specific ansatz used; it can arise in the standard $f(R)$ case as well. The upper bound on the contrary is different since the conformal transformation used in this case is only in the three metric; however, in general, we can say that it is in agreement with the $f(R)$ case studied before since the numbers are not too far apart. Since the $F$ function has the same general form as in the particular case, the projectable version could be studied in the same way, and it led to the same form as in the particular case. Once again, the dS conjecture could be valid for all values of $\bar{R}$, but not independently of it. In this case, we obtained a new lower bound for $\lambda$ that originated an upper bound for $\mu$ and those $\lambda$ became restricted once again to a region not far from $1/3$. This behaviour was shown in Figure~\ref{GeneralLambdasPPr}. The $f(R)$ limit could be studied in this case as well, and we obtained a more restrictive condition on $n$ which could allow for writing the $f$ function as a GR term plus corrections.

In the general case, we were able to study the scenario with a constant Hubble parameter, which was not possible in the particular case. We found that, once again, the dS conjecture and the consistency of the theory led to a restriction on the parameters $\mu$ and $\lambda$. However, it was shown that these restrictions forbid the infrared limit $\mu,\lambda\to1$, and thus the agreement with the dS conjecture and an exponentially expanding universe is exclusive to HL $F(\bar{R})$ theories. Let us point out that as we stated in the Introduction, GR is inconsistent with the dS conjecture when there is an scalar field whose potential obeys the slow roll conditions, and thus when the Hubble parameter is approximately constant. Therefore, it is interesting to see that with a constant Hubble parameter that can describe an inflationary scenario, GR and standard $f(R)$ theories are not consistent with the dS conjecture, whereas HL $F(\bar{R})$ theories are, in fact, consistent.

In summary, our results show that by assuming that the dS conjecture holds, we deduced restrictions on the parameters of the $F(\bar{R})$ theories that are in agreement with the UV behaviour in some cases or represent an exclusive behaviour for these theories in the other. Thus, we can conclude that our results support the idea that the dS conjecture may be encoding key aspects of quantum gravity. We also point out that since these theories break Lorentz invariance in the UV but we are still obtaining consistent results, the dS conjecture is found to be applicable to theories without general covariance a posteriori (as was also considered before in \cite{Trivedi:2021nss}), and thus they may be, in fact, more general than originally thought.

Therefore, we believe that our results support existing ideas that these kinds of theories have very interesting features and deserve to be studied further. In particular, it will be interesting to study the astrophysical implications; for example, we know that projectable HL gravity can provide a term in the equations of motion which behaves as dark matter~\cite{Mukohyama:2009mz}, and we saw that this term appears here as well. Thus, these theories could also lead to new features on this scale. It can also be interesting to explore a way to constrain the parameters of the theories with numerical data in the observational side. Some of this work has been carried out for a particular solution in HL gravity in \cite{Harko:2009qr,Escamilla-Rivera:2020llu} or seeking to constrain the $\lambda$ parameter in \cite{Dutta:2010jh,Nilsson:2019bxv,Czuchry:2023rbi}, which signals a value of $\lambda$ close to one. However, these works are carried out in the context of standard HL theories and could differ when the analysis is performed in the $F(\bar{R})$ theories. Furthermore, even if it is found that $\lambda$ and $\mu$ are close to the infrared value, we can still fulfill the dS conjecture as shown in the present article for the \mbox{general case.}

Finally, let us mention a possible scenario to extend this work in the future. The metric of most interest in building cosmological models is the flat FLRW metric. It is in this metric that the description of an inflationary universe with a scalar field in GR is usually performed and where the dS conjecture is not fulfilled. That is the main reason that we used it throughout this work. However, it would be interesting to use a closed or open FLRW metric or even other anisotropic metrics that will lead to different equations of motion between the different versions of the Ho\v{r}ava--Lifshitz $F(\bar{R})$ theory, and thus they will lead to different scenarios regarding the dS conjecture. This could be used, in principle, to obtain a better sense of the version of the theory that is the most useful using the dS conjecture as a guideline.

\vspace{1cm}
\centerline{\bf Acknowledgments} \vspace{.5cm} D. Mata-Pacheco and L. Zapata would
like to thank CONAHCyT for a grant.


\begin{thebibliography}{99}

\bibitem{Vafa:2005ui}
C.~Vafa,
``The String landscape and the swampland,''
[arXiv:hep-th/0509212 [hep-th]].

\bibitem{Brennan:2017rbf}
T.~D.~Brennan, F.~Carta and C.~Vafa,
``The String Landscape, the Swampland, and the Missing Corner,''
PoS \textbf{TASI2017} (2017), 015
doi:10.22323/1.305.0015
[arXiv:1711.00864 [hep-th]].

\bibitem{Ooguri:2006in}
H.~Ooguri and C.~Vafa,
``On the Geometry of the String Landscape and the Swampland,''
Nucl. Phys. B \textbf{766} (2007), 21-33
doi:10.1016/j.nuclphysb.2006.10.033
[arXiv:hep-th/0605264 [hep-th]].

\bibitem{Palti:2019pca}
E.~Palti,
``The Swampland: Introduction and Review,''
Fortsch. Phys. \textbf{67} (2019) no.6, 1900037
doi:10.1002/prop.201900037
[arXiv:1903.06239 [hep-th]].

\bibitem{vanBeest:2021lhn}
M.~van Beest, J.~Calder\'on-Infante, D.~Mirfendereski and I.~Valenzuela,
``Lectures on the Swampland Program in String Compactifications,''
Phys. Rept. \textbf{989}, 1-50 (2022)
doi:10.1016/j.physrep.2022.09.002
[arXiv:2102.01111 [hep-th]].
%182 citations counted in INSPIRE as of 14 Feb 2023

\bibitem{Grana:2021zvf}
M.~Gra\~na and A.~Herr\'aez,
``The Swampland Conjectures: A Bridge from Quantum Gravity to Particle Physics,''
Universe \textbf{7}, no.8, 273 (2021)
doi:10.3390/universe7080273
[arXiv:2107.00087 [hep-th]].
%62 citations counted in INSPIRE as of 14 Feb 2023

\bibitem{Baumann:2014nda}
D.~Baumann and L.~McAllister,
``Inflation and String Theory,''
Cambridge University Press, 2015,
ISBN 978-1-107-08969-3, 978-1-316-23718-2
doi:10.1017/CBO9781316105733
[arXiv:1404.2601 [hep-th]].

\bibitem{Obied:2018sgi}
G.~Obied, H.~Ooguri, L.~Spodyneiko and C.~Vafa,
``De Sitter Space and the Swampland,''
[arXiv:1806.08362 [hep-th]].


\bibitem{Ooguri:2018wrx}
H.~Ooguri, E.~Palti, G.~Shiu and C.~Vafa,
``Distance and de Sitter Conjectures on the Swampland,''
Phys. Lett. B \textbf{788} (2019), 180-184
doi:10.1016/j.physletb.2018.11.018
[arXiv:1810.05506 [hep-th]].


\bibitem{Andriot:2018wzk}
D.~Andriot,
``On the de Sitter swampland criterion,''
Phys. Lett. B \textbf{785} (2018), 570-573
doi:10.1016/j.physletb.2018.09.022
[arXiv:1806.10999 [hep-th]].

\bibitem{Agrawal:2018own}
P.~Agrawal, G.~Obied, P.~J.~Steinhardt and C.~Vafa,
``On the Cosmological Implications of the String Swampland,''
Phys. Lett. B \textbf{784} (2018), 271-276
doi:10.1016/j.physletb.2018.07.040
[arXiv:1806.09718 [hep-th]].


\bibitem{Roupec:2018mbn}
C.~Roupec and T.~Wrase,
``de Sitter Extrema and the Swampland,''
Fortsch. Phys. \textbf{67} (2019) no.1-2, 1800082
doi:10.1002/prop.201800082
[arXiv:1807.09538 [hep-th]].


\bibitem{Garg:2018reu}
S.~K.~Garg and C.~Krishnan,
``Bounds on Slow Roll and the de Sitter Swampland,''
JHEP \textbf{11} (2019), 075
doi:10.1007/JHEP11(2019)075
[arXiv:1807.05193 [hep-th]].


\bibitem{Ben-Dayan:2018mhe}
I.~Ben-Dayan,
``Draining the Swampland,''
Phys. Rev. D \textbf{99} (2019) no.10, 101301
doi:10.1103/PhysRevD.99.101301
[arXiv:1808.01615 [hep-th]].


\bibitem{Kinney:2018nny}
W.~H.~Kinney, S.~Vagnozzi and L.~Visinelli,
``The zoo plot meets the swampland: mutual (in)consistency of single-field inflation, string conjectures, and cosmological data,''
Class. Quant. Grav. \textbf{36} (2019) no.11, 117001
doi:10.1088/1361-6382/ab1d87
[arXiv:1808.06424 [astro-ph.CO]].

\bibitem{Motaharfar:2018zyb}
M.~Motaharfar, V.~Kamali and R.~O.~Ramos,
``Warm inflation as a way out of the swampland,''
Phys. Rev. D \textbf{99} (2019) no.6, 063513
doi:10.1103/PhysRevD.99.063513
[arXiv:1810.02816 [astro-ph.CO]].


\bibitem{Achucarro:2018vey}
A.~Ach\'ucarro and G.~A.~Palma,
``The string swampland constraints require multi-field inflation,''
JCAP \textbf{02} (2019), 041
doi:10.1088/1475-7516/2019/02/041
[arXiv:1807.04390 [hep-th]].

\bibitem{Artymowski:2019vfy}
M.~Artymowski and I.~Ben-Dayan,
``f(R) and Brans-Dicke Theories and the Swampland,''
JCAP \textbf{05} (2019), 042
doi:10.1088/1475-7516/2019/05/042
[arXiv:1902.02849 [gr-qc]].


\bibitem{Benetti:2019smr}
M.~Benetti, S.~Capozziello and L.~L.~Graef,
``Swampland conjecture in $f(R)$ gravity by the Noether Symmetry Approach,''
Phys. Rev. D \textbf{100} (2019) no.8, 084013
doi:10.1103/PhysRevD.100.084013
[arXiv:1905.05654 [gr-qc]].


\bibitem{Elizalde:2022oej}
E.~Elizalde and M.~Khurshudyan,
``Swampland criteria for $f(R)$ gravity derived with a Gaussian process,''
Eur. Phys. J. C \textbf{82} (2022) no.9, 811
doi:10.1140/epjc/s10052-022-10763-6

\bibitem{Denef:2018etk}
F.~Denef, A.~Hebecker and T.~Wrase,
``de Sitter swampland conjecture and the Higgs potential,''
Phys. Rev. D \textbf{98} (2018) no.8, 086004
doi:10.1103/PhysRevD.98.086004
[arXiv:1807.06581 [hep-th]].

\bibitem{Trivedi:2020wxf}
O.~Trivedi,
``Swampland conjectures and single field inflation in modified cosmological scenarios,''
[arXiv:2008.05474 [hep-th]].


\bibitem{Yi:2018dhl}
Z.~Yi and Y.~Gong,
``Gauss\textendash{}Bonnet Inflation and the String Swampland,''
Universe \textbf{5} (2019) no.9, 200
doi:10.3390/universe5090200
[arXiv:1811.01625 [gr-qc]].

\bibitem{Brahma:2019kch}
S.~Brahma and M.~W.~Hossain,
``Dark energy beyond quintessence: Constraints from the swampland,''
JHEP \textbf{06} (2019), 070
doi:10.1007/JHEP06(2019)070
[arXiv:1902.11014 [hep-th]].

\bibitem{Trivedi:2021nss}
O.~Trivedi, ``Lorentz violating inflation and the swampland,''
Eur. Phys. J. Plus \textbf{137} (2022) no.4, 507
doi:10.1140/epjp/s13360-022-02719-2
[arXiv:2106.03578 [hep-th]].

\bibitem{Horava:2009uw}
P.~Horava,
``Quantum Gravity at a Lifshitz Point,''
Phys. Rev. D \textbf{79} (2009), 084008
doi:10.1103/PhysRevD.79.084008
[arXiv:0901.3775 [hep-th]].

\bibitem{Bertolami:2011ka}
O.~Bertolami and C.~A.~D.~Zarro,
``Ho\v{r}ava-Lifshitz Quantum Cosmology,''
Phys. Rev. D \textbf{84} (2011), 044042
doi:10.1103/PhysRevD.84.044042
[arXiv:1106.0126 [hep-th]].

\bibitem{Sotiriou:2010wn}
T.~P.~Sotiriou,
``Ho\v{r}ava-Lifshitz gravity: a status report,''
J.\ Phys.\ Conf.\ Ser.\  {\bf 283}, 012034 (2011)
doi:10.1088/1742-6596/283/1/012034
[arXiv:1010.3218 [hep-th]].

\bibitem{Kiritsis:2009sh}
E.~Kiritsis and G.~Kofinas,
``Ho\v{r}ava-Lifshitz Cosmology,''
Nucl. Phys. B \textbf{821} (2009), 467-480
doi:10.1016/j.nuclphysb.2009.05.005
[arXiv:0904.1334 [hep-th]].

\bibitem{Mukohyama:2010xz}
S.~Mukohyama,
``Ho\v{r}ava-Lifshitz Cosmology: A Review,''
Class.\ Quant.\ Grav.\  {\bf 27}, 223101 (2010)
doi:10.1088/0264-9381/27/22/223101
[arXiv:1007.5199 [hep-th]].


\bibitem{Calcagni:2009ar}
G.~Calcagni,
``Cosmology of the Lifshitz universe,''
JHEP \textbf{09} (2009), 112
doi:10.1088/1126-6708/2009/09/112
[arXiv:0904.0829 [hep-th]].

\bibitem{Brandenberger:2009yt}
R.~Brandenberger,
``Matter Bounce in Ho\v{r}ava-Lifshitz Cosmology,''
Phys. Rev. D \textbf{80} (2009), 043516
doi:10.1103/PhysRevD.80.043516
[arXiv:0904.2835 [hep-th]].

\bibitem{Czuchry:2009hz}
E.~Czuchry,
``The Phase portrait of a matter bounce in Ho\v{r}ava-Lifshitz cosmology,''
Class. Quant. Grav. \textbf{28} (2011), 085011
doi:10.1088/0264-9381/28/8/085011
[arXiv:0911.3891 [hep-th]].

\bibitem{Mukohyama:2009mz}
S.~Mukohyama,
``Dark matter as integration constant in Ho\v{r}ava-Lifshitz gravity,''
Phys. Rev. D \textbf{80} (2009), 064005
doi:10.1103/PhysRevD.80.064005
[arXiv:0905.3563 [hep-th]].


\bibitem{Zhu:2011yu}
T.~Zhu, F.~W.~Shu, Q.~Wu and A.~Wang,
``General covariant Ho\v{r}ava-Lifshitz gravity without projectability condition and its applications to cosmology,''
Phys. Rev. D \textbf{85} (2012), 044053
doi:10.1103/PhysRevD.85.044053
[arXiv:1110.5106 [hep-th]].

\bibitem{Huang:2012ep}
Y.~Huang, A.~Wang and Q.~Wu,
``Inflation in general covariant theory of gravity,''
JCAP \textbf{10} (2012), 010
doi:10.1088/1475-7516/2012/10/010
[arXiv:1201.4630 [gr-qc]].

\bibitem{Chaichian:2010yi}
M.~Chaichian, S.~Nojiri, S.~D.~Odintsov, M.~Oksanen and A.~Tureanu,
``Modified $F(R)$ Ho\v{r}ava-Lifshitz gravity: a way to accelerating FRW cosmology,''
Class. Quant. Grav. \textbf{27} (2010), 185021
[erratum: Class. Quant. Grav. \textbf{29} (2012), 159501]
doi:10.1088/0264-9381/27/18/185021
[arXiv:1001.4102 [hep-th]].

\bibitem{Elizalde:2010ep}
E.~Elizalde, S.~Nojiri, S.~D.~Odintsov and D.~Saez-Gomez,
``Unifying inflation with dark energy in modified $F(R)$ Ho\v{r}ava-Lifshitz gravity,''
Eur. Phys. J. C \textbf{70} (2010), 351-361
doi:10.1140/epjc/s10052-010-1455-7
[arXiv:1006.3387 [hep-th]].

\bibitem{DeFelice:2010aj}
A.~De Felice and S.~Tsujikawa,
``$f(R)$ theories,''
Living Rev. Rel. \textbf{13} (2010), 3
doi:10.12942/lrr-2010-3
[arXiv:1002.4928 [gr-qc]].


\bibitem{Faraoni:2010pgm}
V.~Faraoni and S.~Capozziello,
``Beyond Einstein Gravity: A Survey of Gravitational Theories for Cosmology and Astrophysics,''
Springer, 2011,
ISBN 978-94-007-0164-9, 978-94-007-0165-6
doi:10.1007/978-94-007-0165-6


\bibitem{Wu:2019xtv}
Q.~Wu and T.~Zhu,
``Inflationary Cosmology with Quantum Gravitational Effects and Swampland Conjectures,''
Commun. Theor. Phys. \textbf{71} (2019) no.9, 1115-1120
doi:10.1088/0253-6102/71/9/1115
[arXiv:1912.05145 [gr-qc]].


\bibitem{Lopez-Revelles:2012xal}
A.~J.~Lopez-Revelles, R.~Myrzakulov and D.~Saez-Gomez,
``Ekpyrotic universes in $F(R)$ Ho\v{r}ava-Lifshitz gravity,''
Phys. Rev. D \textbf{85} (2012), 103521
doi:10.1103/PhysRevD.85.103521
[arXiv:1201.5647 [gr-qc]].

\bibitem{Carloni:2010nx}
S.~Carloni, M.~Chaichian, S.~Nojiri, S.~D.~Odintsov, M.~Oksanen and A.~Tureanu,
``Modified first-order Ho\v{r}ava-Lifshitz gravity: Hamiltonian analysis of the general theory and accelerating FRW cosmology in power-law $F(R)$ model,''
Phys. Rev. D \textbf{82} (2010), 065020
[erratum: Phys. Rev. D \textbf{85} (2012), 129904]
doi:10.1103/PhysRevD.82.065020
[arXiv:1003.3925 [hep-th]].


\bibitem{Chaichian:2010zn}
M.~Chaichian, M.~Oksanen and A.~Tureanu,
``Hamiltonian analysis of non-projectable modified $F(R)$ Ho\v{r}ava-Lifshitz gravity,''
Phys. Lett. B \textbf{693} (2010), 404-414
[erratum: Phys. Lett. B \textbf{713} (2012), 514]
doi:10.1016/j.physletb.2010.08.061
[arXiv:1006.3235 [hep-th]].

\bibitem{Kluson:2011ff}
J.~Kluson,
``Note About Equivalence of $F(\widetilde R)$ and Scalar Tensor Ho\v{r}ava-Lifshitz Gravities,''
Phys. Rev. D \textbf{84} (2011), 104014
doi:10.1103/PhysRevD.84.104014
[arXiv:1107.5660 [hep-th]].


\bibitem{Kluson:2010za}
J.~Kluson, S.~Nojiri, S.~D.~Odintsov and D.~Saez-Gomez,
``$U$(1) Invariant $F(\widetilde{R})$ Ho\v{r}ava-Lifshitz Gravity,''
Eur. Phys. J. C \textbf{71} (2011), 1690
doi:10.1140/epjc/s10052-011-1690-6
[arXiv:1012.0473 [hep-th]].



\bibitem{Harko:2009qr}
T.~Harko, Z.~Kovacs and F.~S.~N.~Lobo,
``Solar system tests of Ho\v{r}ava-Lifshitz gravity,''
Proc. Roy. Soc. Lond. A \textbf{467} (2011), 1390-1407
doi:10.1098/rspa.2010.0477
[arXiv:0908.2874 [gr-qc]].

\bibitem{Escamilla-Rivera:2020llu}
C.~Escamilla-Rivera and H.~Quevedo,
``Quantum signatures from Horava\textendash{}Lifshitz cosmography,''
Class. Quant. Grav. \textbf{38} (2021) no.11, 115009
doi:10.1088/1361-6382/abf66c
[arXiv:2005.08784 [gr-qc]].

\bibitem{Dutta:2010jh}
S.~Dutta and E.~N.~Saridakis,
``Overall observational constraints on the running parameter $\lambda$ of Horava-Lifshitz gravity,''
JCAP \textbf{05} (2010), 013
doi:10.1088/1475-7516/2010/05/013
[arXiv:1002.3373 [hep-th]].

\bibitem{Nilsson:2019bxv}
N.~A.~Nilsson,
``Preferred-frame effects, the $H_0$ tension, and probes of Ho\v{r}ava\textendash{}Lifshitz gravity,''
Eur. Phys. J. Plus \textbf{135} (2020) no.4, 361
doi:10.1140/epjp/s13360-020-00369-w
[arXiv:1910.14414 [gr-qc]].

\bibitem{Czuchry:2023rbi}
E.~Czuchry and N.~A.~Nilsson,
``On the energy flow of $\lambda$ in Ho\v{r}ava-Lifshitz cosmology,''
[arXiv:2304.09766 [astro-ph.CO]].

\end{thebibliography}
\end{document}